\documentclass[aps,prl,showpacs,twocolumn,superscriptaddress,floatfix]{revtex4-2}
\usepackage[utf8]{inputenc}
\newcommand{\figurescale}{1}
\usepackage{verbatim}
\usepackage{amsmath}
\usepackage{mathtools}

\DeclarePairedDelimiterX\braket[2]{\langle}{\rangle}{#1 \delimsize\vert #2}

\usepackage[pdftex]{graphicx}
\usepackage[separate-uncertainty=true]{siunitx}
\DeclareSIUnit{\rpm}{rpm}

\usepackage[colorlinks=true,urlcolor=blue,linkcolor=blue,citecolor=blue]{hyperref}
\usepackage[usenames,dvipsnames]{xcolor}
\usepackage[T1]{fontenc}
\usepackage{placeins}
\usepackage{nicefrac}
\usepackage{braket}
\usepackage{pdfcomment}
\usepackage{xcolor}
\usepackage{braket}
\usepackage[normalem]{ulem}

\usepackage{lineno}

\begin{document}

%
\title{Quantitative electron beam-single atom interactions enabled by sub-20-pm precision targeting}
%
%
%
\author{Kevin~M.~Roccapriore}\email{roccapriorkm@ornl.gov}
\affiliation{Center for Nanophase Materials Sciences, Oak Ridge National Laboratory, Oak Ridge, TN, 37830, USA}
\author{Frances~M.~Ross}
\affiliation{Department of Materials Science and Engineering, Massachusetts Institute of Technology, Cambridge, MA 02139, USA}
\author{Julian~Klein}\email{jpklein@mit.edu}
\affiliation{Department of Materials Science and Engineering, Massachusetts Institute of Technology, Cambridge, MA 02139, USA}
%
%
%
\date\today
%
%
\begin{abstract}
\textbf{
The ability to probe and control matter at the picometer scale is essential for advancing quantum and energy technologies. Scanning transmission electron microscopy offers powerful capabilities for materials analysis and modification, but sample damage, drift, and scan distortions hinder single atom analysis and deterministic manipulation. Materials analysis and modification via electron-solid interactions could be transformed by precise electron delivery to a specified atomic location, maintaining the beam position despite drift, and minimizing collateral dose. Here we develop a fast, low-dose, sub-20-pm precision electron beam positioning technique, "atomic lock-on,” (ALO), which offers the ability to position the beam on a specific atomic column \textit{without} previously irradiating that column. We use this technique to lock onto the same selected atomic location to repeatedly measure its weak electron energy loss signal despite sample drift. Moreover, we quantitatively measure electron beam matter interactions of single atomic events with $\SI{}{\micro\second}$ time resolution. This enables us to observe single atom dynamics such as atomic bistability in the electron microscope, revealing partially bonded atomic configurations and recapture phenomena. We discuss the prospects for high-precision measurements and deterministic control of matter for quantum technologies using electron microscopy.
}
\end{abstract}


%
%
\maketitle
%

Modern aberration-corrected scanning transmission electron microscopy (STEM) underpins our understanding of the structure, chemistry and bonding of materials. It enables us to probe atomic arrangements~\cite{Batson.2002,Erni.2009,Zhou.2013,Krivanek.2014}, measure elemental identity and electronic configuration through spectroscopy~\cite{Krivanek.2010,Suenaga.2010,Zhou.2012,Krivanek.2014}, and control matter at the atomic scale by displacing atoms to trigger material reactions~\cite{Batson2007,Susi2014,Susi2017,Tripathi2018,Dyck.2019}. Controlled nanoscale material modifications are particularly intriguing, as single atomic defects in layered materials can serve as sources of quantum light;~\cite{Montblanch2023} however, their deterministic generation has not yet been achieved. Despite these advances, several major challenges remain. The interaction of high-energy electrons with the atoms in the sample often leads to unwanted sample damage~\cite{Egerton.2004}. Additionally, sample drift and scan distortions during acquisition, even at the Ångström level, complicate signal collection and interpretation~\cite{vonHarrach1995,Jones2013}. Overcoming these challenges would revolutionize our capability to analyze materials through imaging and spectroscopy as well as to drive and quantify electron-solid interactions~\cite{Egerton.2004,GarcadeAbajo2010,Susi2019}, ideally at the single-atom level. 

The remarkable measurements of which STEM is capable, such as electron energy loss spectroscopy (EELS) of individual atom columns, are generally achieved by raster scanning a finely focused beam over a sample area and recording data at each point, which is then presented as a map of intensity or energy~\cite{Ramasse2013,Wang2017}. Repeated raster scanning allows dynamic processes to be studied, including amorphization, recrystallization, phase transformations~\cite{Huang2013,Lin2014,Klein.2022} or the movement~\cite{Isaacson1977,Batson2007,Girit2009,Lee2013,Kotakoski2014,Susi2014,Li2017} or displacement~\cite{Komsa2012,Meyer2012,Kretschmer2020,Speckmann2023} of atoms. However, at the single-atom level, measuring fast dynamics or highly localized signals requires the ability to systematically target a single atomic column, defect, or bond with the beam, \textit{without} dosing nearby locations. This capability has so far been out of reach, but is fundamental to enabling compositional analysis of specific sites and the manipulation of individual atoms or columns of atoms.

\begin{figure*}
	\scalebox{\figurescale}{\includegraphics[width=1\linewidth]{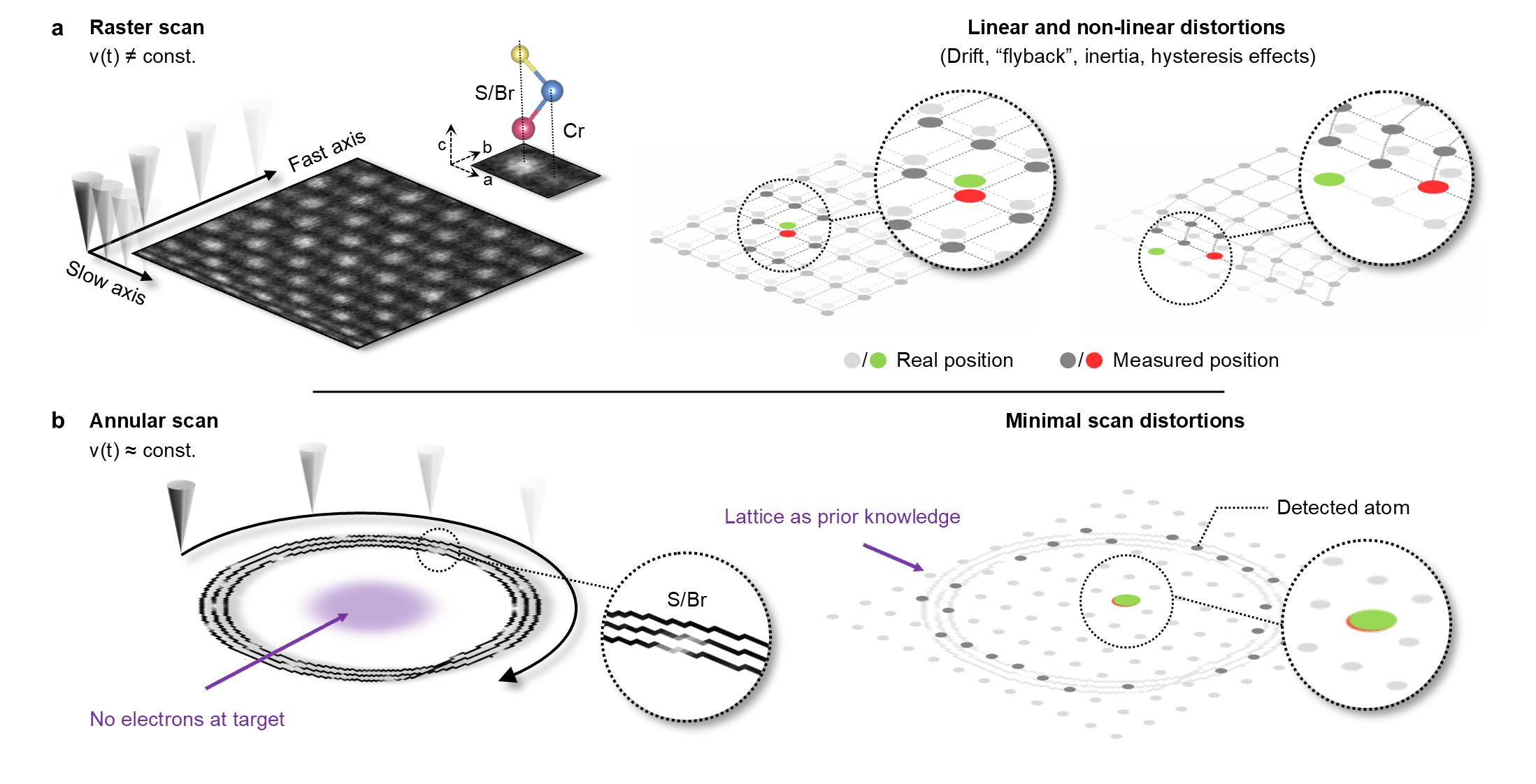}}
	\caption{\label{fig1}
		\textbf{\textit{In situ} positioning inaccuracy due to linear and nonlinear imaging distortions in STEM.}
		\textbf{a}, Conventional raster scan showing the "fast" and "slow" scan axes and a non-uniform electron beam velocity ($v(t) \neq const.$), illustrated with CrSBr. The real atomic lattice $L'(x,y)$ (light gray) or target atom position (green) deviates from the measured lattice $L(x,y)$ (dark gray) or target atom position (red).
        \textbf{b}, Annular scan with a near-constant electron beam velocity ($v(t) \approx const.$). The minimization of scan distortions in the annular scan allows the real atomic lattice to be inferred from sparse atom column information using atomic lock-on because $L'(x,y) \sim L(x,y)$.}
\end{figure*}

\textit{In situ} electron beam positioning has been an unsolved challenge mainly because positional information requires exposing the area of interest to a high electron dose. Moreover, conventional raster scanning limits acquisition speed and exacerbates imaging distortions and sample drift (Fig.~\ref{fig1}a). Sample drift results in a deviation of the measured from the real lattice coordinates, $L'(x,y) \neq L(x,y)$, while, even more critically, the nonlinear distortions prevalent in raster scans~\cite{vonHarrach1995,Jones2013} nontrivially convolve the real lattice coordinates into different measured coordinates $L'(x,y) = f(L(x,y))$, due to the nonuniform beam movement, inertia of the magnetic scan coils at high scan speeds, and abrupt changes in scan direction that take place along the "fast" and "slow" scan axes and during rapid repositioning ("flyback").

These challenges substantially complicate single atom measurements, particularly when performed manually~\cite{Susi2014,Tripathi2018} or (more recently) assisted by deep convolutional neural networks (DCNNs)~\cite{Roccapriore2022,Roccapriore.2024}. Although atom column positions can be extracted using DCNNs in images, they do not reflect the real position of the site of interest due to the accumulation of scan distortions and drift in the collected image (Fig.~\ref{fig1}a). Moreover, since all current approaches for positioning require an image to be obtained, the material experiences a high electron dose that can change or damage it, including at the area of interest. Correcting image distortions can be achieved but it requires a substantial dose; furthermore, since the correction is applied during post-processing steps it is not suitable for \textit{in situ} experiments~\cite{Sang.2014,Ophus.2016}. This combination of challenges strongly motivates the development of an \textit{in situ} electron beam positioning strategy that circumvents the effects of drift, scan-based distortions and excessive irradiation dose.

We address these challenges using a sparse annular scan pattern (Fig.~\ref{fig1}b). This has several advantages. First and most critically, the target region receives no dose. Sparse scanning requires significantly fewer total electrons for lattice reconstruction; therefore, it is fast and minimally invasive for the studied material, preventing unintended structural modifications. Second, the radial symmetry and uniform velocity of the electron beam together substantially reduce the scan distortions~\cite{Sang.2016,Sang.2017}. The sparse annular scan parameters are optimized to collect just enough information to reconstruct the lattice structure when combined with \textit{a priori} structural information of the crystal under study. Using this approach, the measured atomic lattice now matches the real atomic lattice coordinates much more closely, $L'(x,y) \sim L(x,y)$ (Fig.~\ref{fig1}b).

\begin{figure*}
	\scalebox{\figurescale}{\includegraphics[width=1\linewidth]{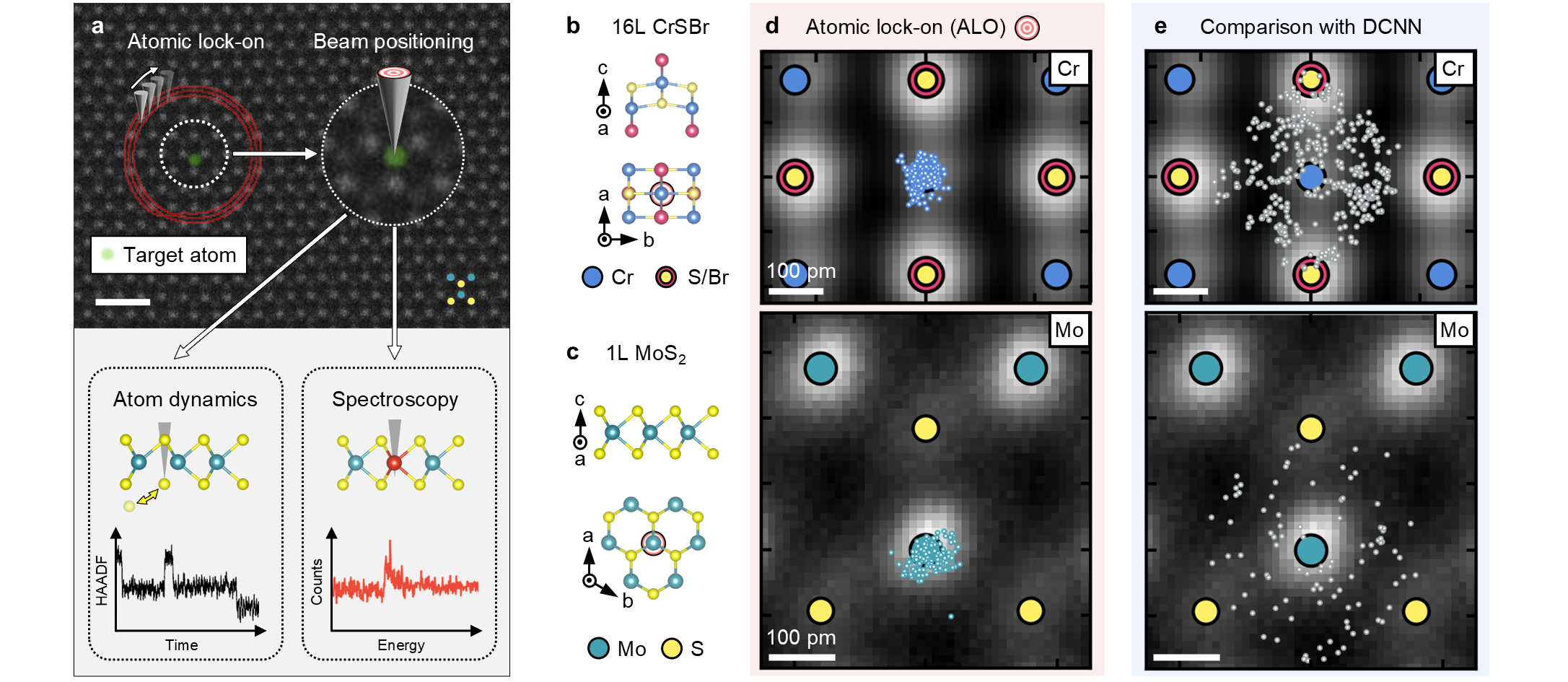}}
	\caption{\label{fig2}
		\textbf{"Atomic lock-on" for few-picometer \textit{in situ} electron-beam targeting in STEM.}
        \textbf{a}, Illustration summarizing use cases, where precision in atomic targeting enables quantitative single-atom dynamics and spectroscopy. Scale bar is $\SI{1}{\nano\meter}$.
        \textbf{b}, Atomic structure of CrSBr. 
        \textbf{c}, Atomic structure of MoS$_2$.
        \textbf{d}, The accuracy of positioning on a Cr atom column over multiple experiments using atomic lock-on, showing a precision of $18 \pm 10 \SI{}{\pico\meter}$ in 16L CrSBr. Targeting the Mo atom in 1L MoS$_2$
        with atomic lock-on shows a precision of $29 \pm 15 \SI{}{\pico\meter}$ in multiple experiments.
        \textbf{e}, Result of multiple positioning experiments using a deep convolutional neural network (DCNN) with Cr as the target atom column in 16L CrSBr and Mo as target atom in a monolayer (1L) of MoS$_2$ both with a precision $> 100 \SI{}{\pico\meter}$.}
\end{figure*}

We refer to this \textit{in situ} electron beam targeting technique as "atomic lock-on" (ALO). We show below that ALO enables positioning of the electron beam on a chosen location of a crystal rapidly, with low dose, and with sub-$\SI{20}{\pico\meter}$ precision. We have developed a workflow that achieves positioning by performing a sparse annular scan followed by a fast lattice reconstruction that makes use of prior knowledge of the target material's crystal structure. Using typical annular scan radii in the range $\sim \SI{1}{\nano\meter}$, we avoid exposing the target area with electrons before the beam is placed accurately on the target, and require only low dose for the annular scan. After describing the workflow, we show high-precision STEM experiments that are enabled by ALO (Fig.~\ref{fig2}a). These include targeted spectroscopic measurements essential for quantitative analytical studies at the single-atom level, the monitoring and control of material modifications such as the generation of single atomic defects, and the quantification of single-atom dynamics accessed with high time resolution through targeted beam positioning. The precise beam placement of ALO reveals single-atom dynamic phenomena that have not previously been resolved using electron microscopy, such as single atom bistability behavior with the highest temporal resolution so far achieved, 10 $\SI{}{\micro\second}$, limited only by the detector and measurement noise. We suggest that positioning the electron beam in STEM using ALO will offer further exciting possibilities and establish a pathway to quantitatively probe and deterministically control matter.


We first demonstrate the effectiveness of ALO for beam positioning and compare it to targeting using a DCNN. We perform two experiments, selecting as target sites a Cr atom column in a thick crystal of CrSBr (16 layers (L), $\sim\SI{13}{\nano\meter}$ thick), and a Mo atom column in a MoS$_2$ monolayer (1L, $\sim\SI{0.8}{\nano\meter}$ thick) (Fig.~\ref{fig2}b and c). We obtain the target position by using either the ALO algorithm or from analyzing a "parent image" using a state-of-the-art DCNN similar to that used in Ref.~\cite{Roccapriore2022,Roccapriore.2024}.

After targeting with either method, we then assess the accuracy with which the site was located. This assessment is done by collecting a small ($\SI{1}{\nano\meter}$ FOV) spiral high-angle annular dark field (HAADF)-STEM image, centered around the beam position, from which the target atom column position is visible (SI Data Fig.~1). The distortions in this small area scan are small and allow us to accurately determine the precision from the atom column position. The offset between the desired location and the center of this small image tells us the precision of the targeting process. We repeat multiple times to obtain a scatter plot of offset values.

Consistently throughout all our experiments using the DCNN, we obtain a spread of beam positions around the Cr target atom in thick CrSBr (Fig.~\ref{fig2}e), extending up to half a unit cell towards an incorrect atom column. This is expected, since the parent image from which positions are predicted shows effects of the accumulated scan distortions and sample drift. In contrast, we repeatably achieve a robust, sub-20 picometer precision in targeting the Cr atom column by using ALO (Fig.~\ref{fig2}d). Even on 1L MoS$_2$, the more experimentally challenging sample, we demonstrate repeated precision of sub-$\SI{30}{\pico\meter}$ in targeting a Mo atom (Fig.~\ref{fig2}d). Multiple attempts using the DCNN again show a broad spread of positions relative to the target site, (Fig.~\ref{fig1}a and b). We attribute the slightly lower precision for ALO in this sample to lower signal-to-noise ratio and geometric lattice inhomogeneities in freely suspended monolayers. SI Table 1 summarizes the advantages of \textit{in situ} beam positioning using ALO.

We therefore conclude that a sparse HAADF-STEM scan is successful in identifying atomic columns and positioning an electron beam on an individual atomic location. The precision of this positioning is limited by thermal lattice vibrations that introduce dynamic displacements of atoms from their equilibrium positions, blurring the atomic potential seen by the beam. Additionally, the finite probe size, delocalization due to inelastic scattering, and long-range Coulomb interactions can further reduce spatial precision. The interaction volume and spatial uncertainty are broadened compared to the accuracy of the algorithm, setting a fundamental upper limit to the positioning precision.

\begin{figure*}
	\scalebox{\figurescale}{\includegraphics[width=1\linewidth]{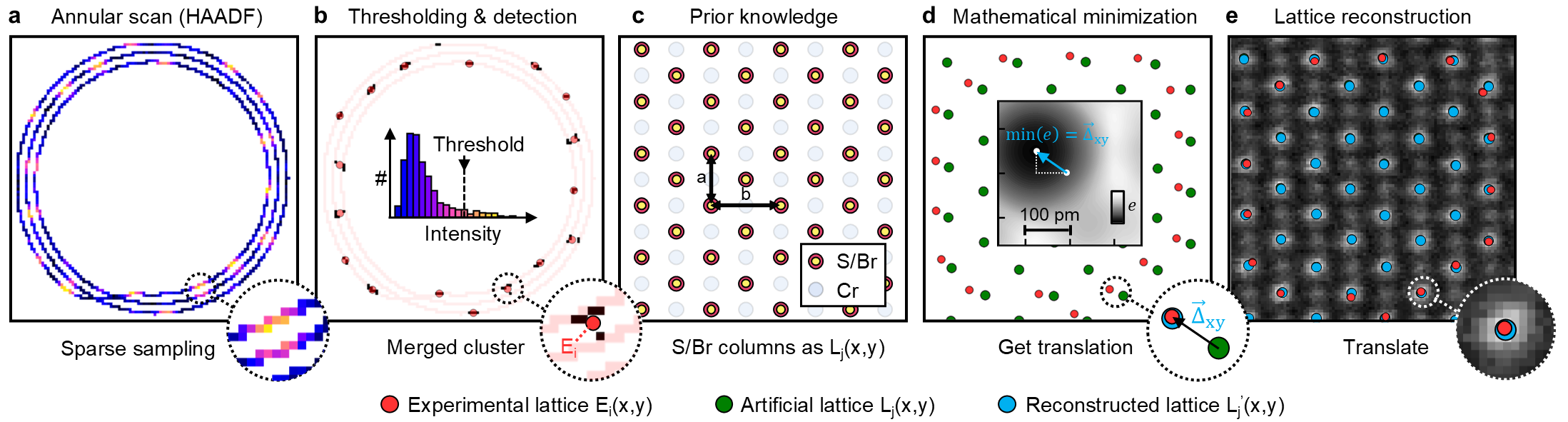}}
	\caption{\label{fig3}
		\textbf{\textit{In situ} lattice reconstruction via atomic lock-on, without prior observation of the target site.}
        \textbf{a}, Annular scan and corresponding HAADF detector signal (blue = low counts, yellow = high counts) showing sparsely scanned S/Br atom columns.
        \textbf{b}, Histogram of HAADF-STEM scan with threshold filtering used to identify pixels that are in S/Br atom columns. Pixel clusters are merged and the center of mass determines the experimental atom position $E_i$.
        \textbf{c}, Prior knowledge is used to construct an artificial lattice $L_j(x,y)$ from lattice parameters (a and b) and rotation angle $\theta$.
        \textbf{d}, Mathematical minimization to obtain the optimal translation vector $\vec{\Delta}_{xy}$ that results in the minimum residual $e$ (inset) between artificial and experimental lattice.
        \textbf{e}, The S/Br sub-lattice is reconstructed by translating the artificial lattice.
		}
\end{figure*}

The ALO scan typically takes $\SI{100}{\milli\second}$ at a beam current of $\SI{20}{\pico\ampere}$, resulting in a low dose of $D = 9.19 \cdot 10^4 e^-$ $\SI{}{\per\angstrom\squared}$ at a dose rate of $\dot{D} = 9.19 \cdot 10^5 e^-$ $\SI{}{\per\angstrom\squared\per\second}$ (see more details in the Methods section). Following the scan, lattice reconstruction requires less than $\SI{10}{\milli\second}$, with additional operating system processes accounting for the remainder of a $\SI{200}{\milli\second}$ "off" time. Throughout this "off" period, the beam is electrostatically blanked to prevent damage to the material. We emphasize that ALO can be performed without prior imaging, as long as the sample is in focus; the lattice can be reconstructed from any scan position.

Figure~\ref{fig3} summarizes the workflow of ALO. Figure~\ref{fig3}a shows the sparse annular scan which consists of three loops and has an outer radius of $\SI{1}{\nano\meter}$ and inner radius of $\SI{0.8}{\nano\meter}$. This is performed (instead of a full-area raster scan) immediately before targeting a desired site, and can be repeated as needed during extended experiments over time intervals that depend on the stability of the microscope, the sample, and other factors, as we discuss below. As the beam on its annular path passes over the atom columns they generate a higher detector output signal: for the example of CrSBr, the sub-lattice of S/Br columns shows up most brightly. We then isolate the pixels associated with this sub-lattice by thresholding the annular scan signal (Fig.~\ref{fig3}b). We merge clusters of pixels dispersed along the circumferential and radial directions to obtain the experimental lattice positions $E_i(x,y)$ of the sparsely measured S/Br sub-lattice. We use atomic lattice constants as \textit{a priori} information (Fig.~\ref{fig3}c) to generate an artificial lattice $L_j(x,y)$ (Fig.~\ref{fig3}d). This is followed by a mathematical minimization to find the best overlap of the artificial lattice with the experimental points via residual minimization to obtain the optimal translation vector $\vec{\Delta}_{xy}$. In the final step, we add $\vec{\Delta}_{xy}$ to the lattice point locations (Fig.~\ref{fig3}e) to obtain the translated lattice $L'_j(x,y)$. Now any position is known by a translation relative to the reconstructed sub-lattice. This allows us to determine the location of any site within the unit cell without directly observing that site with electrons. In this example we reconstruct the S/Br sub-lattice, but the approach is independent of crystal symmetry and we have applied it to mono- and multi-atomic compounds combining different atomic numbers, provided the material is periodic. Despite the presence of crystal distortions and defects in monolayers, we find ALO to be robust, attributed to the detection of multiple atoms in the annular scan. This minimizes the effects of defects and distortions, maintaining a sub-30 pm accuracy (see Fig.~\ref{fig2}d) even in monolayers. The precision provided via ALO consequently enables us to repeatedly target individual atoms with the electron beam \textit{without} previously observing them, even in the sensitive monolayer case.

We systematically optimized the scan parameters for ALO, such as inner and outer radius, number of roundtrips and pixel dwell time, to achieve the highest precision with the least electron dose (SI Data Fig. 2, 3 and 4). This involved performing a grid search simulating ALO and then reducing the dwell time to a level where we reconstruct the lattice with < 50pm precision 100 percent of the time. While ALO works for many choices of inner and outer radii, we found that three roundtrips are necessary for robust positioning. The details of the optimization are discussed in the Supporting Information.

\begin{figure*}
	\scalebox{\figurescale}{\includegraphics[width=0.657\linewidth]{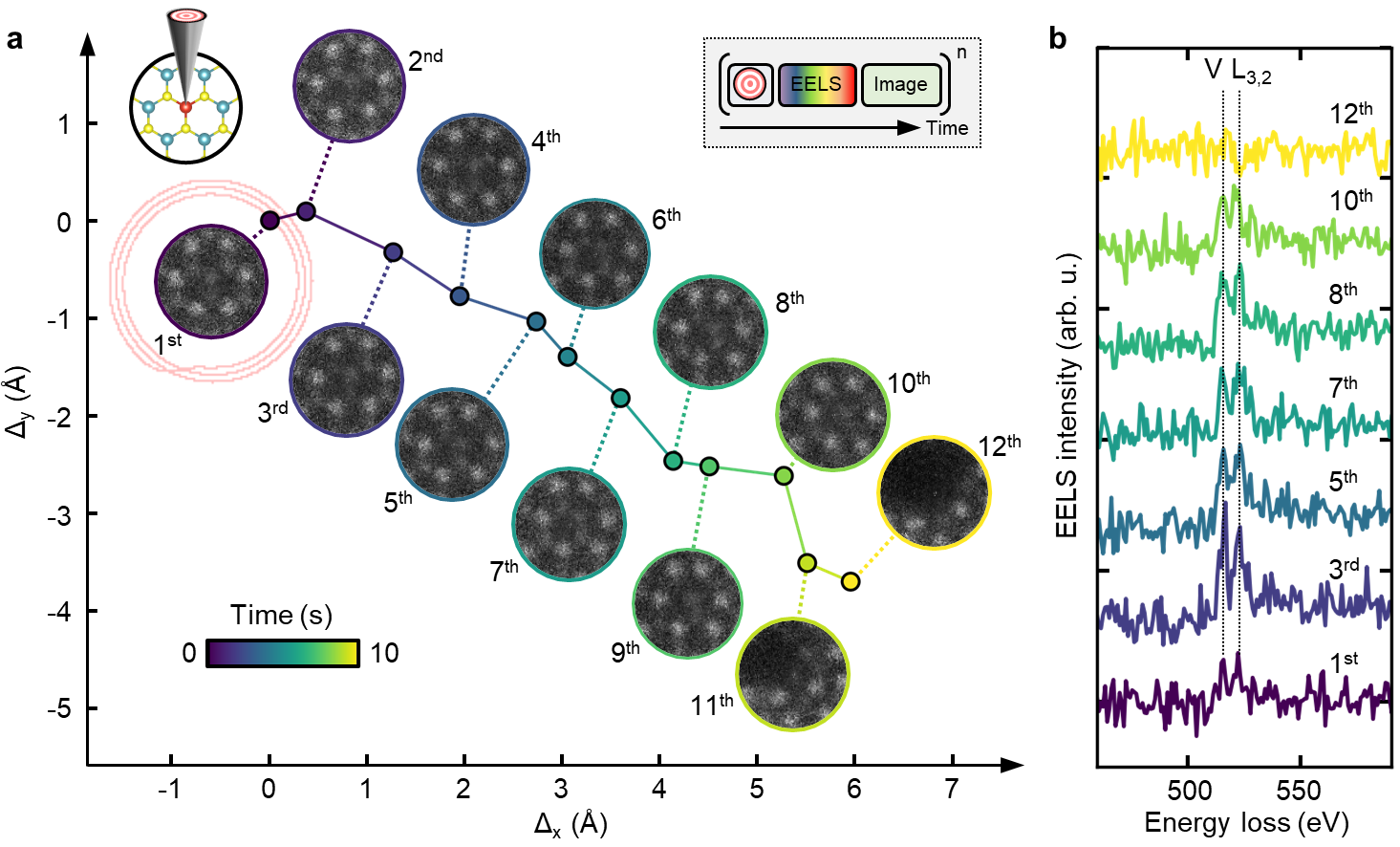}}
	\caption{\label{fig4}
		\textbf{Dynamic targeting and single-shot localized EELS on a V dopant atom in 1L MoS$_2$.}
		\textbf{a}, Time dependent tracking performing 12 consecutive atomic lock-ons on the V dopant atom. The accumulated and compensated drift $\vec{\Delta}_{xy}$ is shown in time alongside collected small HAADF-STEM image. Inset: Automated experimental sequence.
        \textbf{b}, Corresponding single-shot EELS spectra collected subsequent to atomic lock-on showing the L$_{3,2}$ edge of the V dopant atom. The EELS integration time per spectrum was $250 \SI{}{\milli\second}$ at a beam energy of $\SI{60}{\kilo\electronvolt}$ and a beam current of $\SI{20}{\pico\ampere}$ corresponding to $3.12 \cdot 10^7 e^-$ at a dose of $D = 6.2 \cdot 10^7 e^- \SI{}{\per\angstrom\squared}$ and a dose rate of $\dot{D} = 2.5 \cdot 10^8 e^-\SI{}{\per\angstrom\squared\per\second}$.}
\end{figure*}

The most challenging of STEM measurements involve probing a specific location in a sample over an extended time, for example to integrate weak spectroscopic signals or measure single-atom dynamics. We describe below examples of both of these measurements, where the beam is placed on the target with sub-20pm precision and unnecessary dosing of the target area is avoided. Instead, repeated operation of ALO, which can be carried out in an automated sequence, compensates for sample drift to locate, track and measure one specific atom within a particular unit cell. 


By keeping the beam on a single dopant atom for times exceeding 1 second we measure its weak EELS core loss signal. This is shown for a single V dopant atom in 1L MoS$_2$ in Fig.~\ref{fig4}. Quantifying a single dopant atom is a challenging task~\cite{Ramasse2013,Robertson2016}, and therefore ideal for this demonstration. Because our target is a single, non-periodic feature, (a V dopant atom), we first must collect a single parent image. (We emphasize that this step is not necessary if the sample is periodic). We then identify the initial location to be targeted with a pre-trained DCNN~\cite{Roccapriore.2024}. From this point on, ALO is used; we find that the DCNN alone is not sufficiently accurate to provide the correct dopant location. We then carry out repeated operations consisting of ALO at the selected dopant atom followed by collection of a single-shot EEL spectrum from this location with a sampling time of $\SI{830}{\milli\second}$. We repeat the lock-on and spectroscopy 12 times over 10 seconds. This time interval is chosen based on the drift rate of the microscope: it is important to re-center the dopant atom before the drift approaches one unit cell, since the annular scan analysis algorithm can not distinguish between unit cells. The value of $\vec{\Delta}_{xy}$ obtained in each operation is effectively the (compensated) sample drift in time. Figure~\ref{fig4}a shows the values of $\vec{\Delta}_{xy}$ obtained during an experiment while Fig.~\ref{fig4}b shows the EELS spectra obtained after selected ALO operations. Each EELS spectrum show the $L_{3,2}$ edge with expected energies of $\SI{513}{\electronvolt}$ and $\SI{521}{\electronvolt}$. The experiment has therefore successfully tracked the V dopant atom for multiple seconds while compensating for maximum drift rates of $>\SI{1}{\angstrom\per\second}$ while we have also demonstrated extended operation for even higher drift rates of $\SI{2}{\angstrom\per\second}$ (SI Data Fig. 5).

To validate the positioning and observe the environment around the dopant atom, in this experiment we also recorded a small ($\SI{1}{\nano\meter}$ FOV) HAADF-STEM spiral image after each EELS measurement. Acquiring these images is not strictly necessary for the primary measurement, and adds dose to the target area which can influence the local environment. However, the images provide valuable diagnostic information: confirming the stability of the target site during acquisition, monitoring changes in atomic coordination, and identifying transient states. In principle, the dose in such images can be optimized to yield only the essential information, such as the continued presence of the dopant atom. In the images, Fig.~\ref{fig4}a, the V atom appears as a darker feature compared to Mo. It is successfully tracked for over 8 seconds. At this time, the recorded images change, as the accumulated dose has visibly damaged the material. The spectroscopic and imaging data present opportunities for correlation. The changing intensity ratio and line broadening throughout later iterations ($5^{th}$ to $10^{th}$) is suggestive of electronic changes, potentially due to the formation of S vacancy defects visible from lattice distortions. The increasing prevalence of lattice distortions due to repeated measurements have slightly displaced the V dopant atom from the corrected center location of ALO, but is still close enough to the beam location to provide an EELS signal.

The approach in Figure~\ref{fig4} offers the advantage of significantly reducing the electron dose needed to obtain the same information, when compared with grid-based spectrum imaging techniques (SI Data Fig. 6). In techniques where a spectrum is collected from each point and those spectra that came from the area of interest are selected afterwards, there is a greater possibility of inducing structural modifications during acquisition. An intriguing possibility suggested by the approach in Figure~\ref{fig4} is to collect spectroscopy data not only at individual atoms but also at bonds or other specific points within the lattice (SI Data Fig. 6). We finally note that this experiment is scalable in the sense that for any material and spectroscopic measurement, the atom can be tracked \textit{without directly observing it} for as long a time as needed via repeated lock-ons, terminating when sufficient data is collected or the sample becomes damaged.

\begin{figure*}
	\scalebox{\figurescale}{\includegraphics[width=1\linewidth]{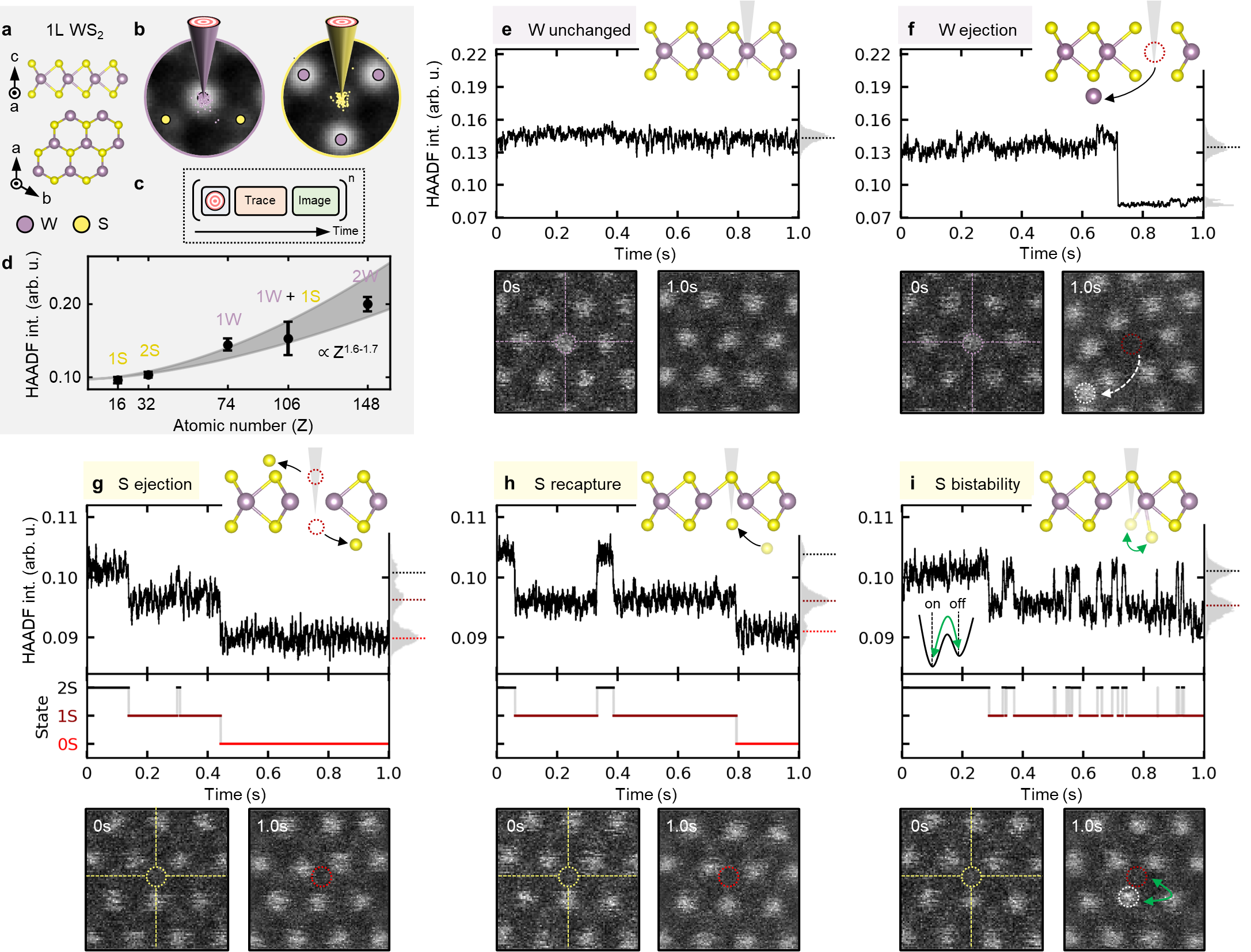}}
	\caption{\label{fig5}
		\textbf{Single-atom dynamics with sub-millisecond time resolution in 1L WS$_2$.}
		\textbf{a}, 1L WS$_2$ crystal structure.
        \textbf{b}, Atomic lock-on targeting of the W atom and 2S atom column with $28 \pm 18 \SI{}{\pico\meter}$ ad $27 \pm 17 \SI{}{\pico\meter}$ precision, respectively.
        \textbf{c}, Experimental workflow.
        \textbf{d}, HAADF intensity as a function of atomic number Z. The dashed grey line is a model $\propto Z^{1.6-1.7}$ to the data.
        Atomic lock-on targeting of \textbf{e-f}, W atom column and \textbf{g-i}, 2S atom column for $\SI{1}{\second}$ and corresponding HAADF-STEM image before and after experiment. The corresponding HAADF intensity histogram depicts different structural states. A moving average of $\SI{3}{\milli\second}$ was applied to the time traces. A beam energy of $\SI{60}{\kilo\electronvolt}$ and current $\SI{20}{\pico\ampere}$ were used corresponding to $1.25 \cdot 10^8 e^-$ for $\SI{1}{\second}$ at a dose of $D = 6.2 \cdot 10^7 e^- \SI{}{\per\angstrom\squared}$ and a dose rate of $\dot{D}=2.5 \cdot 10^8 e^-\SI{}{\per\angstrom\squared\per\second}$.}
\end{figure*}


The ability to position the electron beam precisely on one specific atom then record a detector signal as a function of time offers a pathway to study the interaction between electrons and individual atoms. In Fig.~\ref{fig5}a we show such an experiment performed on monolayer thick WS$_2$ where the HAADF signal is recorded in time. For both the W and the 2S sites, lock-on obtains consistent sub-$\SI{30}{\pico\meter}$ precision (Fig.~\ref{fig5}b and SI Data Fig. 7). In the automated experiment (Fig.~\ref{fig5}c), we therefore first perform ALO to position the electron beam on either a W or 2S site. We then monitor the HAADF intensity with a fast sampling frequency of $\SI{100}{\kilo\hertz}$ ($\SI{10}{\micro\second}$) for a total time of $\SI{1}{\second}$. Due to the relationship between HAADF signal and atomic number, we interpret the intensity in terms of the atomic configuration at the target site as a function of time (Fig.~\ref{fig5}d). As in the spectroscopy experiment, this data acquisition is followed by the collection of a small ($\SI{1}{\nano\meter}$ FOV) diagnostic HAADF-STEM image of the final structure, centered on the target site. We carry out 50 such targeting experiments automatically and sequentially on different W and 2S sites to obtain statistics regarding the range of behavior. A single experiment takes $\sim \SI{1.5}{\second}$ resulting in a total experimental acquisition time for each type of target site of $\sim \SI{75}{\second}$. The ability to repeat the measurement quickly makes this approach highly scalable for rapidly collecting statistical data.

We show a selection of recurring experimental behaviors at the W atom site in Fig.~\ref{fig5}, e and f and at the 2S atom site in Fig.~\ref{fig5}, g to i. Additional data is shown in (SI Data Fig. 8 and Fig. 9). When targeting W, 25\% of experiments show no appreciable change in the time-dependent HAADF intensity (Fig.~\ref{fig5}e). The final image is also unchanged. In these experiments, positioning the electron beam on a W atom has therefore not led to rearrangement, which can be attributed to the high displacement threshold energy~\cite{Komsa2012}. However, 4\% of the experiments do show displacement, visible both from the step-like drop in HAADF intensity (Fig.~\ref{fig5}e at $\sim\SI{0.7}{\second}$) and from the final image, in which the W atom is displaced and its final location at a neighboring 2S site. Other experiments (SI Data Fig. 8) show step-like changes in the HAADF intensity that may even be an increase over the initial intensity. We attribute these to diffusion of adatoms across the target location. The remaining experiments (making up the majority, 60\%) exhibit intensity fluctuations that can be identified as arising from distortions due to defects induced in the 2S sites in proximity to the W atom (SI Data Fig. 8). 

Targeting the 2S  site, Fig.~\ref{fig5}, g to i, is even more interesting because of the lower displacement threshold energy of S~\cite{Komsa2012} that results in the formation of a mono-sulfur vacancy $V_{1S}$. The experiments consistently (28\%) show a signature of S atom ejection and $V_{1S}$ formation in the form of a step-like decrease in the HAADF intensity (Fig.~\ref{fig5}, g to i and SI Data Fig. 9). In many instances, this is followed by a second drop, directly suggesting the ejection of a second S atom and formation of a di-sulfur vacancy $V_{2S}$. 

More unexpected is that 2/3 of the experiments exhibiting an S atom ejection show the \textit{recapture} of S atoms after the initial assumed displacement (Fig.~\ref{fig5}h at $\sim\SI{0.35}{\second}$ and SI Data Fig. 9). In the other 1/3, we observe random telegraph noise (RTN) (Fig.~\ref{fig5}k at $>\SI{0.28}{\second}$) between the HAADF intensity levels associated with the 2S and 1S configurations. This suggests fast movement of the S atom between the targeted 2S atom column ('on' state) and a proximal site ('off' state). We interpret the recapture and RTN as indications that the S atom remains partially bonded to the lattice. In this scenario, the energy of an electron is sufficient to drive the transition between the two metastable states (2S$\leftrightarrow$1S), thereby effectively acting as a bistable atomic system. The shortest duration we can resolve the S atom in the 2S state is $<\SI{1}{\milli\second}$ ($<1.2 \cdot 10^5 e^-$) suggesting even faster time dynamics. Our interpretation is further supported by examination of the final HAADF image, which shows increased intensity in one of the three nearest neighbor W sites. Based on the excellent agreement of the intensity at this location with the expected atomic number of 1W + 1S (Fig.~\ref{fig5}d), we tentatively attribute this intensity to the S atom residing in a potential 'off' state.  


We have shown two quantitative high-precision STEM experiments: measuring the weak spectroscopic signal of a single dopant atom with acquisition times of up to a second, even in the presence of sample drift, and characterizing the dynamics of individual atoms with sub-ms time resolution. These experiments were enabled by precision beam placement through a rapid algorithm that does not dose the area of interest and can be repeated as needed to locate and track a single atomic site. We showed that ALO can be applied to both thick and thin crystals and we find it particularly promising for challenging beam-sensitive and atomically thin materials. 

The ability to integrate ALO within virtually any STEM workflow, as used in our measurements of beam-induced displacement, offers exciting possibilities for collection of large numbers of datasets. This enables a statistical quantification of phenomena of interest, which can otherwise be a limitation of traditional STEM observations. We also anticipate that minor algorithm modifications may extend ALO to moderately disordered systems or those with specific structural features (such as twin boundaries), increasing the range of materials that can be examined. For example, ALO could look for multiple characteristic arrangements in a sparse annular scan to accurately determine a phase boundary or interface region and perform positioning on either side. In the experiments described so far, the target location was within the annular scan but the beam can be positioned far from the scan area, with accuracy depending on intrinsic imperfections in the sample (due to defects, strain fields or sample bending) and the magnetic scan coil hardware and voltage to picometer calibration. Moreover, while ALO enables precise positioning within the xy-plane, future advancements may include mitigating drift in the out-of-plane (z) direction.

The high temporal resolution combined with single-atom sensitivity and real-time single-event detection lays the foundation for algorithmic schemes to drive and control atomic modifications. An immediate outcome could be the deterministic generation of single-atomic defects in materials, which is essential for scaling quantum technologies~\cite{Montblanch2023}. We further anticipate other uses of ALO for $\textit{in situ}$ STEM studies to observe or control dynamic processes, such as the early stages of nucleation and growth of atomic assemblies. Here, the combination of knowledge of the lattice coordinates and precise positioning of the beam allows single-atom events to be monitored. The measured intensity provides a natural avenue for end-point detection. After a displacement event is triggered, as measured by the change in ADF signal, the beam can be repositioned to another atom column. This would be useful if we need to continue acquisition, for example, of EELS signals, or move a series of atoms in minimal time. ALO is applicable generally to crystalline materials with high Z contrast specimens most readily amenable, but we anticipate its use even for low-Z materials such as graphene and hexagonal boron nitride, for example if we wish to generate defects. Atomic manipulation by ALO can be synchronized with a range of analytical measurement techniques such as EELS, energy-dispersive X-ray spectroscopy or even 4D-STEM, to perform multimodal studies that obtain chemical and electronic information with minimal disturbance to the local environment. Combining simultaneous data collection with the ability to move between equivalent locations as a function of the evolving atomic configuration could enable acquisition of weak spectroscopic signals, for example for a single S vacancy in MoS$_2$ or WS$_2$, challenging due to the low signal to noise ratio. Such measurements could benefit experiments where external sample parameters (current, doping density, electric field, light, temperature or even the gas environment) are controlled, for example in catalytic processes. Precise positioning might also be useful to locally inject electrons to trigger single-defect cathodoluminescence or electron beam-induced current measurements. 

We finally suggest that ALO will open a pathway towards a more systematic understanding of elastic and inelastic scattering mechanisms of individual atoms or columns, their dependence on beam energy and dose rate, and their evolution over different time and energy scales. We anticipate that site-specific monitoring will shed light on the complex kinetics of beam-driven structural phase changes~\cite{Huang2013,Lin2014,Klein.2022} or the effects of beam-generated secondary electrons. We believe this will contribute to verifying and improving existing classical and quantum mechanical formulations of electron-beam matter interactions~\cite{GarcadeAbajo2010,Yoshimura2023}, refining element-specific displacement cross sections~\cite{Komsa2012,Meyer2012,Kretschmer2020,Speckmann2023}, and studying how the electron beam drives defect formation, migration, and annihilation with individual atom precision.

%
%

\section{Acknowledgements}
J.K. and F.M.R. acknowledge funding through the NSF Trailblazer Award Number 2421694 and DOE-BES Award Number DE-SC0025387. This research was supported by the Center for Nanophase Materials Sciences (CNMS), which is a US Department of Energy, Office of Science User Facility at Oak Ridge National Laboratory. The authors acknowledge the MIT SuperCloud and Lincoln Laboratory Supercomputing Center for providing (HPC, database, consultation) resources that have contributed to the research results reported within this article. We gratefully acknowledge Zdenek Sofer for providing CrSBr bulk material and Joshua Robinson for providing V doped MoS$_2$ and Kai Xiao for WS$_2$ materials.

\section{Author contributions}
J.K. conceptualized the project, developed and tested atomic lock-on, K.M.R. and J.K. implemented atomic lock-on and conducted STEM imaging, J.K. prepared the samples, J.K. and K.M.R. analyzed the experimental data, K.M.R., F.M.R and J.K designed the experiments and discussed the results. J.K. wrote the manuscript with input from all co-authors.

%

\section{Methods}

\subsection{Sample fabrication}
Bulk CrSBr flakes were exfoliated using the Scotch tape method onto SiO$_2$/Si substrates. Flake thickness was determined by atomic force microscopy and optical phase contrast. Selected flakes were transferred onto S/TEM compatible sample grids using cellulose acetate butyrate (CAB) as polymer handle. After transfer, CAB was dissolved in acetone and the S/TEM grids were rinsed in isopropanol prior to critical point drying. MoS$_2$ and WS$_2$ were MOCVD and CVD grown and then transferred to S/TEM grids using PMMA and water transfer.

\subsection{Scanning transmission electron microscopy}
A Nion UltraSTEM operating at either 60 or 200 kV, a nominal probe current of 20 pA, and 32 mrad semiconvergence angle was utilized for the CrSBr, MoS$_2$ and WS$_2$ atomic lock-on studies. The HAADF collection angle was 80-200 mrad.

For STEM-EELS, a Nion Iris spectrometer and direct electron detector (Dectris ELA) were used with a dispersion of $\SI{0.90404}{\electronvolt}$ / channel.

Custom scan trajectories were performed using a National Instruments (NI) multifunction I/O field programmable gate array (FPGA, USB-7856) connected to the external scan input and HAADF output using NI terminal block SCB-68A. This platform operates on LabVIEW but commands are accessible via Python directly from the microscope user interface, Nion Swift, which allows to change, e.g., annular spiral scan parameters.~\cite{Sang2016}

An operating field of view (FOV) of $\SI{16}{\nano\meter}$ was used for all experiments. At larger fields of view (e.g., $> \SI{100}{\nano\meter}$), voltage noise on the scan coils can begin to limit positioning precision. On our system (total scan range of $\pm\SI{2.5}{\volt}$), a $100 \times 100$ pixel scan over a $\SI{2}{\nano\meter}$ annular ALO scan within a $\SI{100}{\nano\meter}$ FOV corresponds to voltage steps of approximately $\SI{1}{\milli\volt}$. The scan controller used exhibits an root mean square (RMS) voltage noise of $\sim\SI{250}{\micro\volt}$ (DC to \SI{1}{\mega\hertz}), which translates to a spatial noise of roughly $\SI{5}{\pico\meter}$ at this scale. Based on this, we estimate that fields of view up to $\sim\SI{200}{\nano\meter}$ remain compatible with sub-\SI{20}{\pico\meter} targeting precision using this controller.

\subsection{Calculation of electron dose and dose rate for atomic lock-on}

We consider the total trajectory covered during the three loops of the annular scan, and the area of the focused electron beam with a $FWHM =  \SI{0.8}{\angstrom}$. To calculate the dose rate $\dot{D}$ and dose $D$, we start by determining the number of electrons reaching the surface over the total exposure time. For the ALO scan time of $\SI{100}{\milli\second}$ and for a beam current of $\SI{20}{\pico\ampere}$, the number of electrons, $N_e$, is calculated as $ N_e = \frac{I \cdot t}{q} = \frac{20 \cdot 10^{-12} \, \text{A} \cdot 0.1 \, \text{s}}{1.602 \cdot 10^{-19} \, \text{C}} \approx 1.248 \cdot 10^7 \, \text{electrons}$. The beam makes three loops with radii $\SI{10}{\angstrom}$, \SI{9}{\angstrom}, and \SI{8}{\angstrom}. The area covered for one donut is $A_{donut}= \pi \cdot (R^2 - r^2)$ with $R = \text{radius} + \frac{\text{FWHM}}{2}$ and $r = \text{radius} - \frac{\text{FWHM}}{2}$. The total area covered by the beam is $A_{\text{total}} = A_{donut,1} + A_{donut,2} + A_{donut,3} = \SI{135.72}{\angstrom\squared}$. The dose rate is given by $\dot{D} = \frac{1.248 \cdot 10^7}{\SI{135.72}{\angstrom\squared}\cdot \SI{0.1}{\second}} = 9.19 \cdot 10^5 e^-$ $\SI{}{\per\angstrom\squared\per\second}$ and the corresponding dose for a $\SI{100}{\milli\second}$ ALO is given by $D = 9.19 \cdot 10^4 e^-$ $\SI{}{\per\angstrom\squared}$. 

During the time to process the atomic lock-on scan ($\sim \SI{200}{\milli\second}$), the beam is electrostatically blanked to avoid dosing the material. We note that the atomic lock-on algorithm itself is fast with <10ms. The atomic lock-on scan can be performed without obtaining a parent image of the area of interest as it obtains lattice information during the atomic lock-on scan, considering a focused condition.

\subsection{Calculation of electron dose for a single spot exposure}

The total electron dose for a time-dependent single spot exposure is calculated by the total number of electrons in time divided by the effective area of the focused electron beam. We consider a focused electron beam with a diameter equal to the $FWHM =  \SI{0.8}{\angstrom}$ resulting in an area of $A = \pi \left( \frac{\text{FWHM}}{2} \right)^2 = \SI{0.5}{\angstrom\squared}$. The total number of electrons that are deposited to the spot during exposure are $N_e = I \cdot t \cdot \frac{1}{e}$. For a beam current of $\SI{20}{\pico\ampere}$ and an exposure time of $\SI{1}{\second}$ we determine the total number of electrons of $N_e = 1.25 \cdot 10^8 \, e^-$. Dividing the total number of electrons $N$ by the effective area of the focused beam $A$ we obtain an electron dose of $2.5 \cdot 10^{8} e^- \SI{}{\per\angstrom\squared}$. For an exposure time of $\SI{250}{\milli\second}$ we obtain a total number of electrons of $N_e = 3.1 \cdot 10^7 e^-$ at an electron dose of $D = 6.2 \cdot 10^7 e^- \SI{}{\per\angstrom\squared}$ at a dose rate of $\dot{D} = \frac{1.248 \cdot 10^8}{\SI{0.5}{\angstrom\squared}\cdot \SI{1}{\second}} = 2.5 \cdot 10^8 e^-\SI{}{\per\angstrom\squared\per\second}$.

\subsection{Data post processing}
We performed no post-processing of collected HAADF-STEM images; only raw data were used and shown throughout this manuscript.

For the EELS data, a power law background  was fitted to the region 420 - 900 eV and was subtracted from each spectrum.

\subsection{Neural network architecture}
Atom detection used in this manuscript for CrSBr, MoS$_2$, and WS$_2$ and utilized the ensemble learning approach described in Refs.~\cite{Roccapriore2022,Ghosh2021}. This consists of DCNNs based on the UNET architecture~\cite{Ronneberger2015}. V dopant detection was based on a previously described learning approach in Ref.~\cite{Roccapriore.2024}.

\subsection{Copyright notice}

This manuscript has been authored by UT-Battelle, LLC, under contract DE-AC05-00OR22725 with the US Department of Energy (DOE). The US government retains and the publisher, by accepting the article for publication, acknowledges that the US government retains a nonexclusive, paid-up, irrevocable, worldwide license to publish or reproduce the published form of this manuscript, or allow others to do so, for US government purposes. DOE will provide public access to these results of federally sponsored research in accordance with the DOE Public Access Plan (https://www.energy.gov/doe-public-access-plan).

\subsection{Data availability}

The data that support the findings of this study are available from the corresponding author upon reasonable request.

\subsection{Competing interests}

The authors are inventors on a filed patent application related to this work.

%
%

\bibliographystyle{naturemag}
\bibliography{full}

\end{document}


\setcounter{figure}{0} 
\renewcommand{\thefigure}{S\arabic{figure}}

\renewcommand{\thetable}{S\arabic{table}}

\fontsize{11pt}{13pt}\selectfont

\makeatletter
\renewcommand\@make@capt@title[2]{%
    \@ifx@empty\float@link{\@firstofone}{\expandafter\href\expandafter{\float@link}}%
    \fontsize{11pt}{13pt}\selectfont\textbf{#1}\@caption@fignum@sep#2
}
\renewcommand\figurename{Figure}
\makeatother

%
\title{Supplemental Information: \\ Quantitative electron beam-single atom interactions enabled by sub-20-pm precision targeting}
%
%
%
\author{Kevin~M.~Roccapriore}\email{roccapriorkm@ornl.gov}
\affiliation{Center for Nanophase Materials Sciences, Oak Ridge National Laboratory, Oak Ridge, TN, 37830, USA}
\author{Frances~M.~Ross}
\affiliation{Department of Materials Science and Engineering, Massachusetts Institute of Technology, Cambridge, MA 02139, USA}
\author{Julian~Klein}\email{jpklein@mit.edu}
\affiliation{Department of Materials Science and Engineering, Massachusetts Institute of Technology, Cambridge, MA 02139, USA}
%
%
\date{\today}
%
%
\maketitle
%
%
\tableofcontents

\newpage

\section{1. Atomic Lock-On and Deep Convolutional Neural Network (DCNN) Targeting Precision}

\begin{figure*}[!ht]
	\scalebox{\figurescale}{\includegraphics[width=0.8\linewidth]{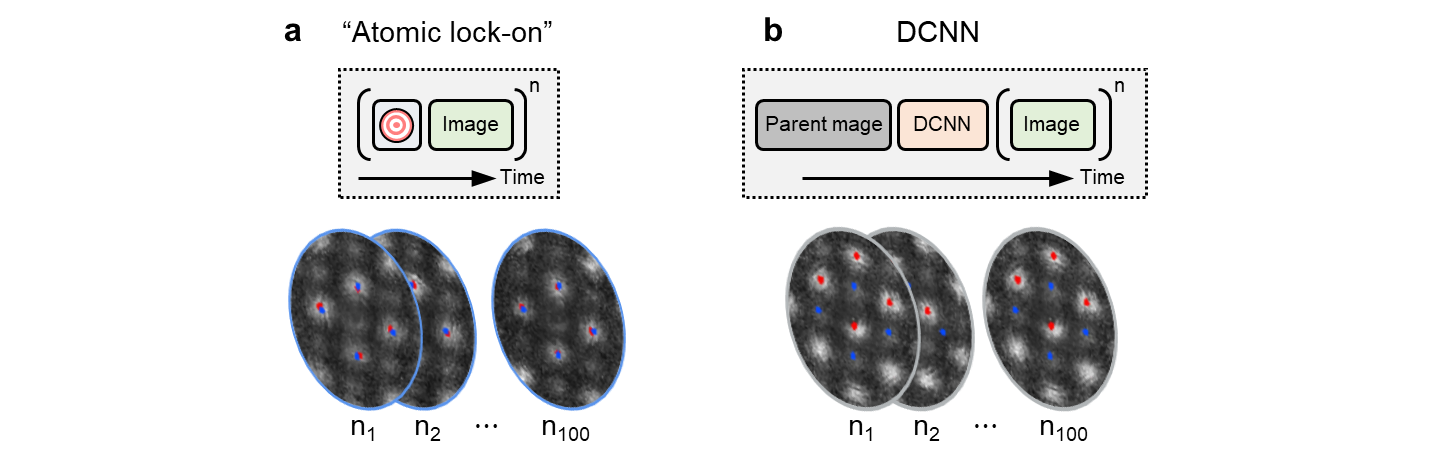}}
	\renewcommand{\figurename}{Figure}
    \caption{\label{figALODCNN}
		\textbf{Experimental comparison between atomic lock-on and a DCNN.}
		\textbf{a}, HAADF-STEM spiral images (1nm FOV) after ALO targeting of the Cr atom column. Red dots are the four nearest neighbor fitted S/Br atom columns which are used to obtain the offset of the Cr atom position from the center of the spiral image. The blue dots are the ideal positions of the S+Br atom columns for perfect targeting.
        \textbf{b}, HAADF-STEM spiral images (1nm FOV) after DCNN targeting of the Cr atom column.
		}
\end{figure*}

\textbf{Figure~\ref{figALODCNN}a ,b} depict how we determine and compare the targeting precision between ALO and using a deep convolutional neural network (DCNN). For each approach, we obtain the position of the target site (e.g., an atom species) and collect a spiral HAADF-STEM image (1nm FOV) centered on the assumed target position. Experimentally, to test the DCNN, we integrate this procedure into an \textit{in situ} workflow that first takes an overview 'parent image' at low magnification (FOV = $\SI{16}{\nano\meter}$) from which atomic column positions are predicted using a DCNN trained on MoS$_2$, WS$_2$ or CrSBr. In the example of CrSBr, in a subsequent step, a subset of detected Cr atom columns are used as targets, and 25 spiral images are collected. This process is repeated four times to obtain a total of 100 spiral images. To verify the precision, after we perform ALO, we collect a small spiral image with an FOV of $\SI{1}{\nano\meter}$ and high pixel density ($\SI{10}{\pico\meter}$/pixel), from which we determine the offset of the Cr atom column from the center of the image by fitting the four nearest neighbor S/Br atom columns. We repeatedly perform ALOs at different locations after each lattice reconstruction. We fit the next nearest-neighbor S/Br atom columns from which we determine the offset of the atom column from the center (\textbf{Fig.~\ref{figALODCNN}a, b}). The reported targeting precision refers to the statistical accuracy with which the electron probe can be positioned at a desired location, considering system stability and drift correction. This is distinct from the interaction spread, which is determined by the physical size of the focused electron probe (about 80 pm FWHM). While targeting precision defines how reproducibly we can position the beam, the interaction spread describes the spatial region over which the electron beam interacts with the sample. Both parameters are critical as precise targeting ensures control over beam placement, while the probe size sets the fundamental spatial resolution limit for interactions such as atomic excitation or displacement.

\begin{table}[htbp]
\centering
\renewcommand{\arraystretch}{1.5}
\setlength{\tabcolsep}{8pt}
\begin{tabular}{|p{4cm}|p{4cm}|p{4cm}|p{3cm}|}
\hline
\textbf{Feature} & \textbf{Current work: Atomic lock-on (ALO) technique (in situ)} & \textbf{Previous work: DCNN-based approach (in situ)} & \textbf{General image precision (ex situ)} \\ \hline

1. Ability to measure the position of an atom & - & - & few-pm \cite{Yankovich.2014} \\ \hline

2. Ability to place the beam on an atom & \textbf{Accurate <20pm} & \textbf{Inaccurate >100pm} & - \\ \hline

3. Electron dose to atom of interest & Atom \textbf{remains undosed} until measurement & Atom \textbf{has been dosed} prior to measurement & - \\ \hline

4. Image requirement & No pre-acquired image necessary (works blind) & Requires pre-acquired (raster-)scanned image & - \\ \hline

5. Distortion and drift compensation & Actively compensates for distortion and drift through continuous acceleration in annular scanning & Ignores distortion and drift (based on static, pre-acquired image) & - \\ \hline

6. Atom position source & Positions derived directly from sparse annular (spiral) beam motion, ensuring real-time accuracy & Positions determined via DCNN analysis of a (raster-)scanned image & - \\ \hline

7. Dosing in region of interest (ROI) & Minimal, targeted dosing only to essential subregions, minimizing unnecessary exposure & All atoms within ROI receive electron dose & - \\ \hline

8. Capability for Time-Resolved Studies & Enables time-resolved studies by preserving the atom of interest, avoiding cumulative exposure & - & Not suitable for tracking single atoms over time due to continuous dosing \\ \hline

\end{tabular}
\caption{Comparison between general \textit{ex situ} atom position precision and \textit{in situ} ALO and DCNN based beam positioning.}
\label{tab:comparison}
\end{table}

\textbf{Table~\ref{tab:comparison}} summarizes the advantages of in situ electron beam positioning of ALO compared to DCNN based positioning. This contrasts with the precision of obtaining the atom position ex situ from a collected image in post-processing. Unlike DCNN, which requires a pre-acquired raster-scanned image and is unable to compensate for distortion or drift, ALO actively compensates through continuous beam acceleration, sparse and fast annular scanning. Importantly, no pre-acquired image is needed. ALO derives atom positions directly from sparse annular scans, ensuring real-time accuracy while preserving the atom of interest by avoiding unnecessary electron dosing prior to measurement. By targeting only essential subregions for dosing, ALO minimizes cumulative exposure, making it suitable for time-dependent studies and enabling minimally invasive observations of single atoms over time. In comparison, ex situ determination of an atom position from a static image gives a precision of a few picometer~\cite{Yankovich.2014}.

\section{2. Grid search optimization for 16L CrSBr and 1L MoS$_2$}

\begin{figure*}[!ht]
	\scalebox{\figurescale}{\includegraphics[width=1\linewidth]{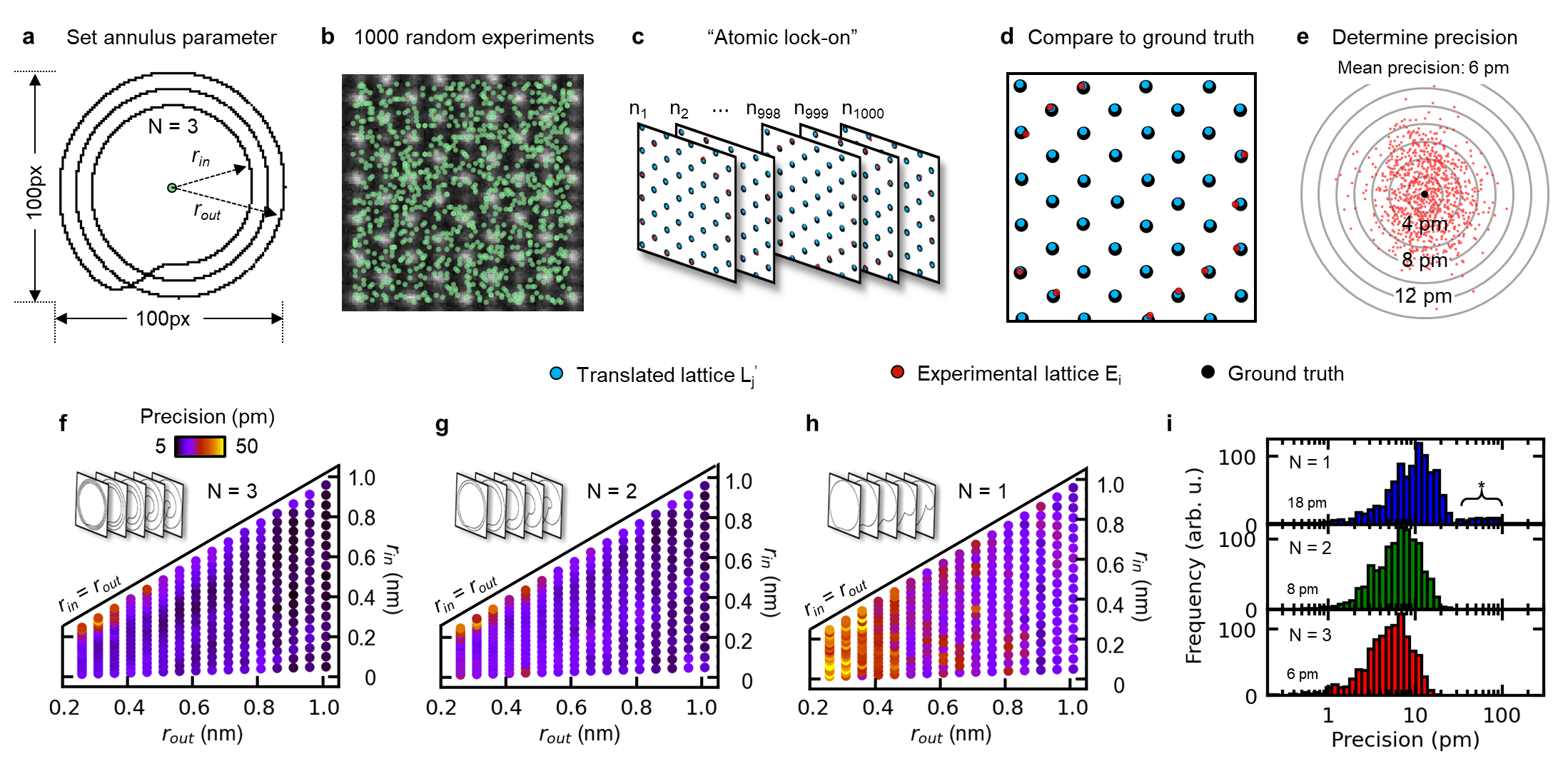}}
    \renewcommand{\figurename}{Figure}
	\caption{\label{figCSBgridsearch}
		\textbf{Determining optimal annular scan parameters using grid search simulation for 16L CrSBr.}
		\textbf{a}, Annular scan pattern, defined by three parameters, the outer radius $r_{out}$, inner radius $r_{in}$ and number of loops $N$ here $N = 3$. Pixel density is $(100px)^2$ and is kept constant.
        \textbf{b}, 1000 random coordinates to statistically evaluate performance of the annular scan at different coordinates.
        \textbf{c}, 1000 ALO simulations at 1000 random coordinates.
        \textbf{d}, Comparison of the reconstructed lattice $L'_j$ with the ground truth lattice positions obtained form a DCNN and 2D Gaussian refinement for obtaining the offset coordinate.
        \textbf{e}, Distribution of offset coordinates with root mean square of $\sim 5 \SI{}{\pico\meter}$ using $r_{out} = \SI{1}{\nano\meter}$, $r_{in} = \SI{0.8}{\nano\meter}$ and $N = 3$.
        \textbf{f-h}, Grid search simulating a total of 912 annular scan shapes for $N = 3$, $N = 2$, and $N =1$, respectively. Each point originates from 1000 simulations.
        \textbf{i}, Histogram showing the simulated precision for $r_{in} = \SI{0.8}{\nano\meter}$ and outer radius $r_{out} = \SI{1}{\nano\meter}$. For each histogram, 1000 random positions were simulated. A single loop ($N = 1$) exhibits failures (highlighted with the *) of ALO absent for $N>1$.
		}
\end{figure*}

\begin{figure*}[!ht]
	\scalebox{\figurescale}{\includegraphics[width=1\linewidth]{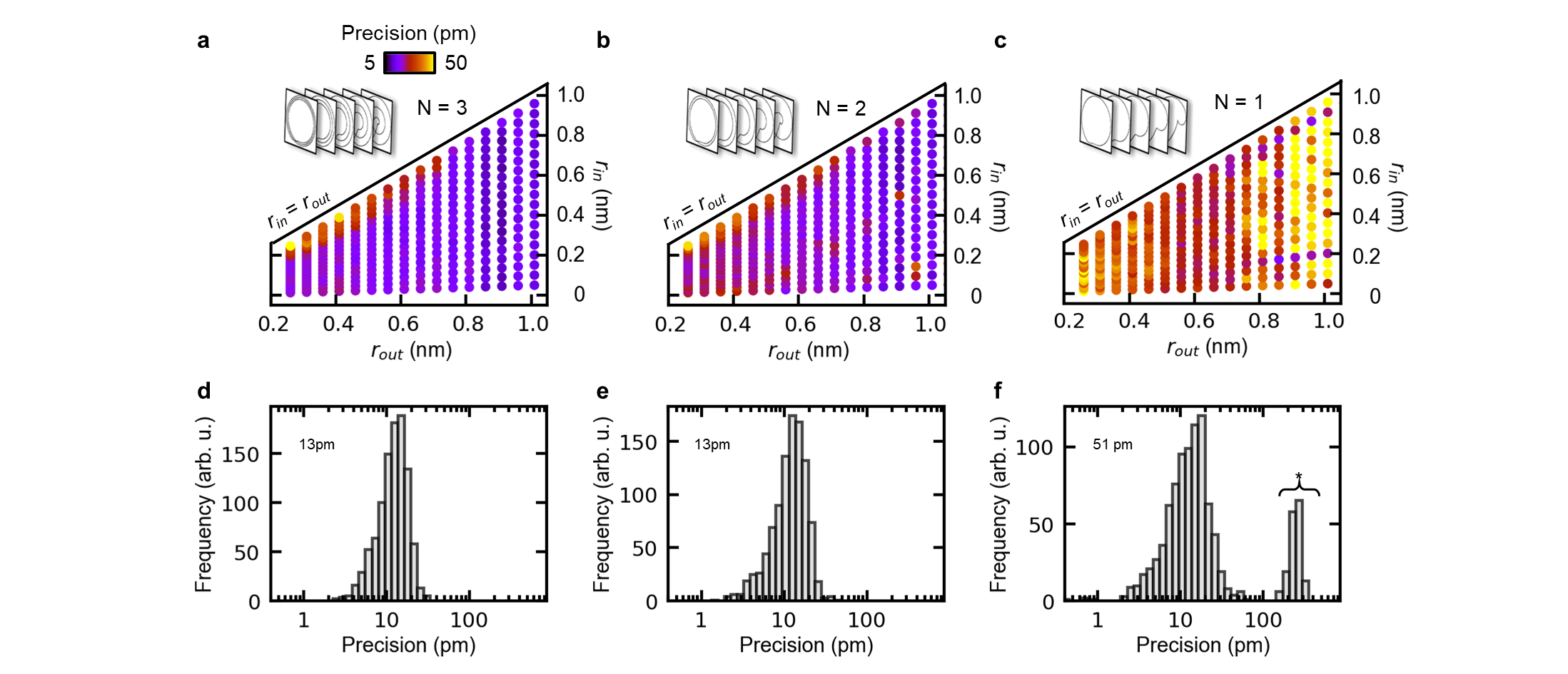}}
    \renewcommand{\figurename}{Figure}
	\caption{\label{figMoS2gridsearch}
		\textbf{Grid search optimization for 1L MoS$_2$}
		\textbf{a-c}, Grid search simulating a total of 912 annular scan shapes for $N = 3$, $N = 2$, and $N =1$, respectively. Each point represents the precision obtained from 1000 random experiments. The precision is shown as a function of inner $r_{in}$ and outer radius $r_{out}$ for different number of loops $N$ of the annulus. 
        \textbf{d-f}, Histogram showing the simulated precision for $r_{in} = \SI{0.7}{\nano\meter}$ and outer radius $r_{out} = \SI{1}{\nano\meter}$. For each histogram, 1000 random positions were simulated. A single loop ($N=1$) exhibits failures (highlighted with the *) of ALO absent for $N>1$.
		}
\end{figure*}

We perform a grid search optimization to obtain the optimal annular scan shape. Our motivation is to determine the annular scan shape with the highest precision and the absence of failures during repeated operations. Our reasoning for this optimization is that the ideal annulus depends on the crystal symmetry and lattice parameters. For the grid search, we consider two different materials, multilayer (16) CrSBr (\textbf{Fig.~\ref{figCSBgridsearch}} and monolayer (1L) MoS$_2$ (\textbf{Fig.~\ref{figMoS2gridsearch}}). We reduce the annulus to three main experimental parameters. In other words, the number of loops $N$, outer radius $r_{out}$ and inner radius $r_{in}$ of the loops (\textbf{Fig.~\ref{figCSBgridsearch}a}). To obtain the optimal annulus shape, a grid search was performed. For this, we vary $N$, $r_{out}$ and $r_{in}$. For each configuration, we generate 1000 random positions, where the annulus is centered on each of these positions (\textbf{Fig.~\ref{figCSBgridsearch}b}) and perform ALO to obtain the translated lattice positions $L_j$ (\textbf{Fig.~\ref{figCSBgridsearch}c}). After each ALO, we compare $L_j$ to the ground truth, which is provided by our HAADF-STEM image (\textbf{Fig.~\ref{figCSBgridsearch}d}). We use a DCNN to predict the atom column positions that are refined in subsequent steps by performing 2D Gaussian fits. From a comparison with the ground truth, we obtain the offset vector, which is our measure of precision. The offset obtained with respect to the ground-truth lattice are shown in \textbf{Fig.~\ref{figCSBgridsearch}e}. We obtain a mean precision of $\sim\SI{5}{\pico\meter}$. This value is comparable to the Gaussian fitting error of the atom columns ($\sim\SI{2}{\pico\meter}$), reflecting the high effectiveness and precision of ALO. 

We now perform a grid search for different numbers of loops $N = [1, 2, 3]$ while varying $r_{out}$ and $r_{in}$ (\textbf{Fig.~\ref{figCSBgridsearch}f, g}) to simulate the mean precision for 304 annular scan shapes for each $N$. We obtain the best results for $N = 3$ for different combinations of $r_{out}$ and $r_{in}$. This suggests a fully robust approach and provides flexibility in choosing experimental values depending on the application. Moreover, we find that $N = 2$ is still robust, albeit with slightly lower precision, whereas a single loop $N = 1$ not only provides reduced precision but also exhibits failures, as highlighted in \textbf{Fig.~\ref{figCSBgridsearch}i}. Therefore, for our experiments, we use $N = 3$ as the ideal trade-off between the highest precision and lowest dose that performs best in challenging experimental environments. For CrSBr, we determine the ideal optimal annular parameters as $r_{out} = \SI{1}{\nano\meter}$ and $r_{in} = \SI{0.8}{\nano\meter}$.

Similar to the optimal annulus used for CrSBr, we also optimize the scan parameters of the annular scan for 1L MoS\textsubscript{2} performing a grid search optimization (\textbf{Fig.~\ref{figMoS2gridsearch}}). We use an experimentally collected HAADF-STEM image for our simulation and define random positions for every set of scan parameters to obtain statistics on the precision. For each of the random positions we perform ALO and determine the lattice parameters based on the obtained translation. We compare this lattice to the ground truth that we obtain from applying a DCNN to the image in combination with a 2D Gaussian atom position refinement step. From the statistics for each set of parameters we obtain a histogram and a precision value. We vary the inner $r_{\text{in}}$ and outer $r_{\text{out}}$ annulus radius and the number of loops as shown in \textbf{Fig.~\ref{figMoS2gridsearch}a-c}. We observe sets of parameters that are favorable to obtain a consistently high precision. We further compare three histograms with 1000 random positions for ALO with $r_{\text{in}} = 0.7$~nm and $r_{\text{out}} = 1$~nm and different numbers of loops (\textbf{Fig.~\ref{figMoS2gridsearch}d-f}). We find that a single loop exhibits failures while for $N > 1$ we obtain a mean precision of $\sim 13$~pm and robust operation.

\section{3. Additional Details of Atomic Lock-On}

\subsection{3.1. Thresholding Condition}

To isolate the pixels associated with the sub-lattice of interest, we threshold the HAADF-STEM annular scan. There are two options to threshold: either using a defined percentage of all pixels based on the expected coverage of pixels of a target atom column in the sparse scan (geometrical argument), or by using the mean intensity of the target atom column of interest. This can be obtained either by manual input or by using a DCNN to detect atom columns in an image.

Here, we applied a DCNN and obtained the atom column intensity distribution from all the atom columns within a HAADF-STEM image with a field of view (FOV) of 16~nm. With respect to this distribution, and in the specific example of CrSBr where we detect the S/Br sub-lattice, we set typical threshold values in the range of 90-120\% with respect to the overall distribution of S/Br columns, depending on the electron dose used for the scan.

\begin{figure*}[!ht]
	\scalebox{\figurescale}{\includegraphics[width=0.7\linewidth]{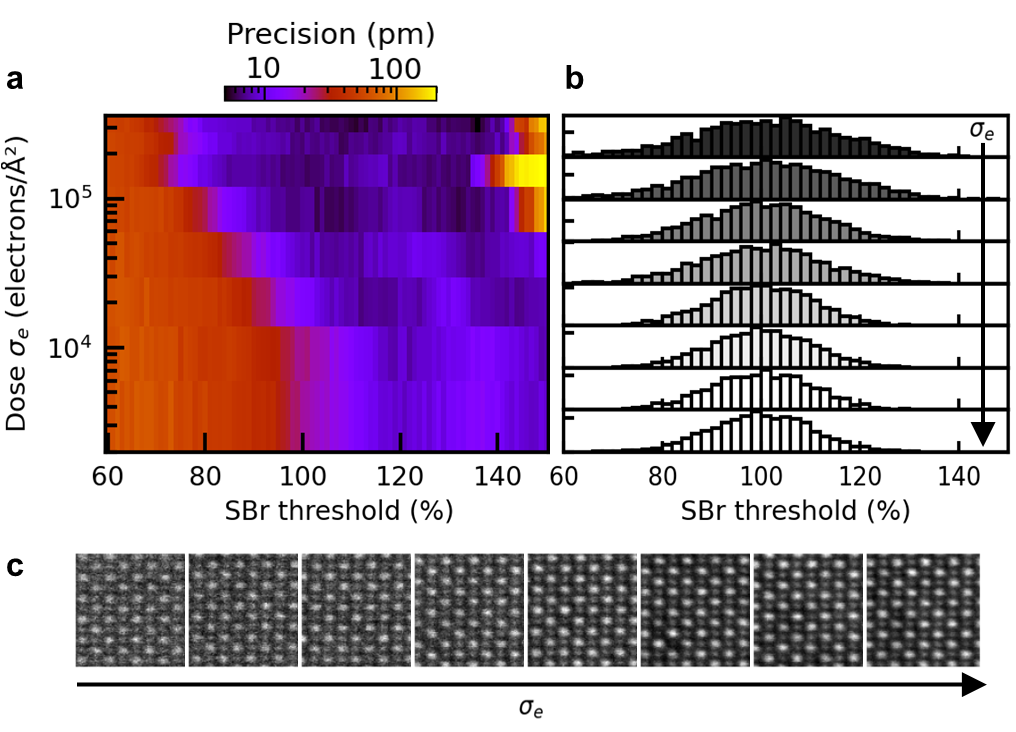}}
    \renewcommand{\figurename}{Figure}
	\caption{\label{fig_noise}
		\textbf{Atomic lock-on precision for different electron doses.}
        \textbf{a}, Simulated mean precision as a function of electron dose $\sigma_e$ and the percentage of the atom column threshold distribution from a DCNN exemplified with the S/Br atom columns in 16L CrSBr.
        \textbf{b}, Intensity distribution of S/Br atom columns with respect to mean intensity obtained from a DCNN.
        \textbf{c}, HAADF-STEM images of CrSBr collected for increasing electron dosages.
		}
\end{figure*}

The threshold value (or threshold window) depends on the material and noise conditions of the data, such as the pixel size and pixel dwell time. We quantify the precision of ALO based on the set threshold level (\textbf{Fig.~\ref{fig_noise}}). For this, we collect experimental HAADF-STEM images at varying electron doses $\sigma_e$ and perform 1000 ALOs at random positions to obtain the corresponding mean precision while varying the threshold value (\textbf{Fig.~\ref{fig_noise}a}). As expected, we find that the onset for obtaining the best precision moves to lower threshold values for increased doses, owing to the narrower atom column intensity distribution (\textbf{Fig.~\ref{fig_noise}b}).

We find that values around 120\% are robust for all doses. Moreover, the precision remains high even for very low electron dosages and can be further decreased, which reduces the number of electrons deposited on the sample and the time required to perform the annular scan. This reduces collateral damage and allows for even higher sampling rates for repeatedly using ALO.

\subsection{3.2. A Priori Information for Lattice Reconstruction}

To perform the lattice reconstruction, an artificial sub-lattice is generated, which is optimized onto the experimental lattice obtained from the thresholding. To generate the sub-lattice, lattice parameters such as lattice vectors, orientation, and rotation are required as input. This information can either be obtained from user input, from detecting the reflections in the fast Fourier transform (FFT) of a HAADF-STEM image, or by using a DCNN that predicts the lattice in a collected HAADF-STEM image. In this study, we successfully tested all three options.

In the next step, we perform a mathematical minimization to determine the best overlap of the artificial lattice positions with the experimental lattice positions. For this, we apply a translation vector $\vec{\Delta}_{xy}$ to all lattice positions. To calculate the residual $e$ for a given translation, we translate the lattice positions as follows

\begin{equation}
    L'_j = L_j + \vec{\Delta}_{xy} \quad \text{for all } j
\end{equation}

where $L'_j$ represents the translated lattice positions. Next, we calculate the distance for each experimental position $E_i$ to the closest translated lattice point $L'_j$ using the Euclidean distance

\begin{equation}
    D_i = \min_j \left( \sqrt{(E_{ix} - L'_{jx})^2 + (E_{iy} - L'_{jy})^2} \right)
\end{equation}

where $E_{ix}$ and $E_{iy}$ are the $x$ and $y$ coordinates of the experimental position $E_i$, and $L'_{jx}$ and $L'_{jy}$ are the $x$ and $y$ coordinates of the translated lattice point $L'_j$. The residual is obtained as

\begin{equation}
    e = \sum_i D_i^2
\end{equation}

The optimal translation vector $\vec{\Delta}_{xy}$ is defined by

\begin{equation}
    \min(e) = \vec{\Delta}_{xy}
\end{equation}

\subsection{3.3. Relative Targeting}

In the final step, we add the translation vector to the artificial lattice positions to obtain the actual lattice positions $L'_j$. In the example of CrSBr, ALO provides the lattice information of the S/Br sub-lattice. To target a specific column in CrSBr, we first select a target site, for example, the Cr atom column. This is straightforward, as we can add a two-dimensional (2D) translation vector with respect to the reconstructed S/Br sub-lattice

\begin{equation}
    \vec{a} = \left(0, \frac{a}{2}\right), \quad \vec{b} = \left(\frac{b}{2}, 0\right)
\end{equation}

for a crystal rotation of $\theta = 0^\circ$. For a finite rotation angle $\theta$, we obtain the rotated coordinates

\begin{equation}
    \vec{a}' = R(\theta) \cdot \vec{a}, \quad \vec{b}' = R(\theta) \cdot \vec{b}
\end{equation}

with the rotation matrix defined as

\begin{equation}
    R(\theta) = 
    \begin{pmatrix}
    \cos\theta & -\sin\theta \\
    \sin\theta & \cos\theta
    \end{pmatrix}
\end{equation}

After rotation, the coordinates of the crystal vectors become

\begin{equation}
    \vec{a}' = \left( -\frac{a}{2} \sin\theta, \frac{a}{2} \cos\theta \right), \quad
    \vec{b}' = \left( \frac{b}{2} \cos\theta, \frac{b}{2} \sin\theta \right)
\end{equation}

\section{4. Automated and dynamic drift compensation.}

\begin{figure*}[!ht]
	\scalebox{\figurescale}{\includegraphics[width=0.8\linewidth]{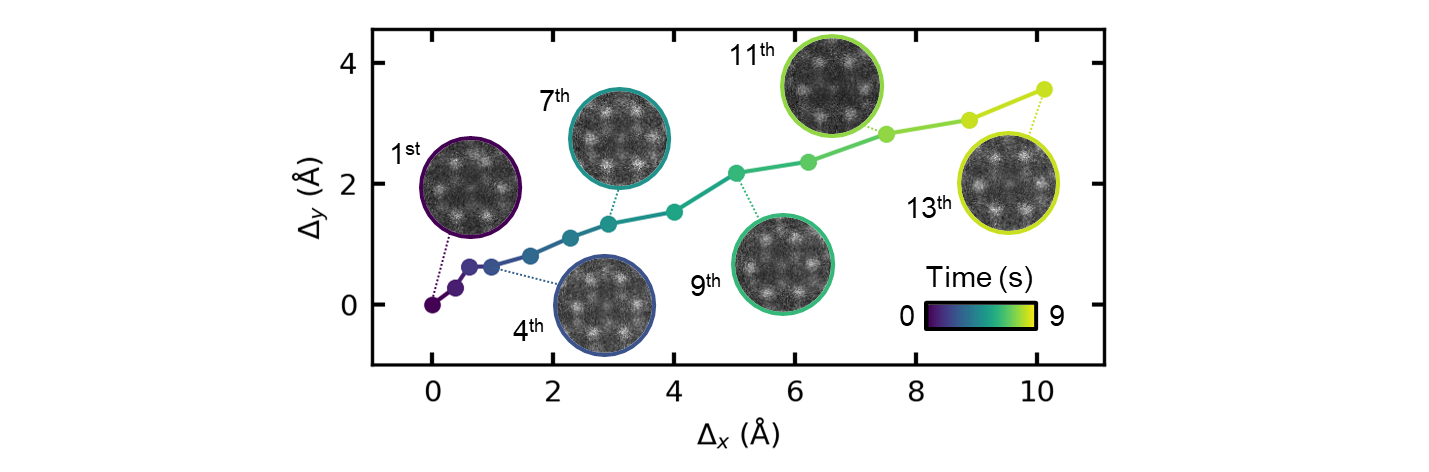}}
    \renewcommand{\figurename}{Figure}
	\caption{\label{fig_S_time}
		\textbf{Automated tracking of a single V dopant atom in 1L WS$_2$.}
		Repeated ALO on a single V dopant atom in 1L WS$_2$. ALO compensates a maximum drift rate of $\SI{2}{\angstrom\per\second}$.
		}
\end{figure*}

\begin{figure*}[!ht]
	\scalebox{\figurescale}{\includegraphics[width=0.8\linewidth]{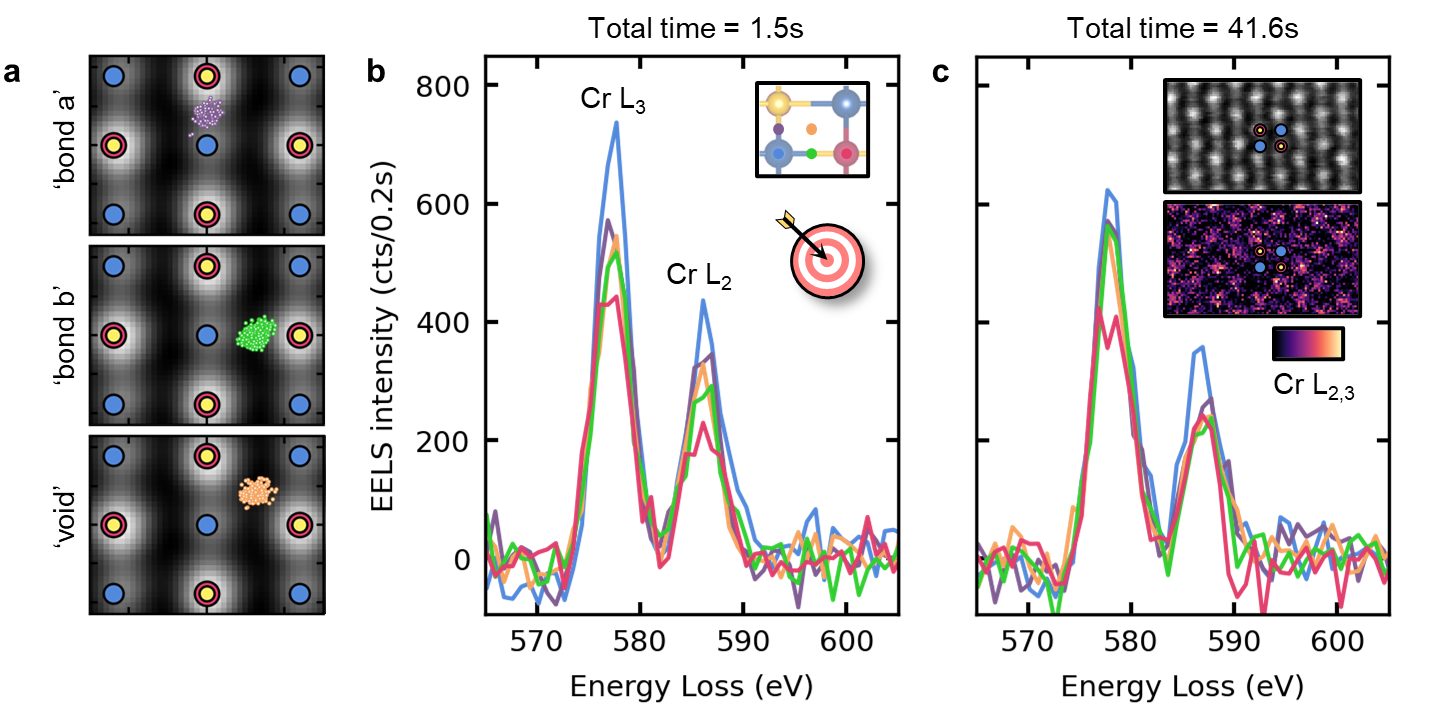}}
    \renewcommand{\figurename}{Figure}
	\caption{\label{figEELSCSB}
		\textbf{Automated single-shot and targeted EELS on atoms and bonds in CrSBr}
        \textbf{a}, ALO targeting precision of the 'a bond', 'b bond' and 'void' in CrSBr with $20 \pm 10 \SI{}{\pico\meter}$, $19 \pm 12 \SI{}{\pico\meter}$, and $20 \pm 10 \SI{}{\pico\meter}$, respectively. 
        \textbf{b}, Targeted single-shot EELS acquired on five target sites (see inset) of the Cr $L_{3, 2}$ edge. Each spectrum is background subtracted to highlight the relative edge intensity. The integration time was $\SI{200}{\milli\second}$ for each spectrum with a beam energy of $\SI{100}{\kilo\electronvolt}$ and a beam current of $\SI{20}{\pico\ampere}$.
        \textbf{c}, Regular raster scan spectral image with a total integration time of $\SI{41.6}{\second}$. 20 individual spectra from each target site are summed to reach the same total dwell time as in single-shot experiments in \textbf{b}. Inset: Simultaneously acquired HAADF-STEM, and an EELS 2D map of the Cr L$_{3,2}$ edge integrated from $\SI{570}{\electronvolt}-\SI{590}{\electronvolt}$ after background subtraction. A beam energy of $\SI{100}{\kilo\electronvolt}$ and a beam current of $\SI{20}{\pico\ampere}$ were used.
		}
\end{figure*}

\section{5. Dose-Effectiveness of Atomic Lock-On.}

Besides spectroscopically probing atoms, reconstructing the lattice by ALO allows sub-atomic placement with the ability to target bonds or other high symmetry sites in the crystal. We demonstrate this with 16L CrSBr by targeting the center of the bond along the a- and the b-direction and the 'void' (\textbf{Fig.~\ref{figEELSCSB}c}). This can have applications where particular beam-driven reactions are governed by beam placement. Verifying the high targeting precision we continue performing single-shot EELS focusing on the Cr $L_{3,2}$ edge as it provides the strongest signal contrast, plus Cr atoms reside in their own atom column. \textbf{Figure~\ref{figEELSCSB}d} shows spectra for five target sites with an integration time of $\SI{200}{\milli\second}$ each. As expected, the Cr column shows the highest intensity and the S/Br the lowest with a contrast of almost 50\%. The presence of Cr signal throughout all positions is from scattering due to the specimen thickness. The total measurement time to collect all five spectra is $\SI{1}{\second}$ and only $\SI{0.5}{\second}$ for performing all ALOs.

We furthermore emphasize the dose-effectiveness of targeted spectroscopic measurements by comparing with a regular pixel-by-pixel ($85$x$49$) spectral image EELS with a size of ($2$x$1$ $\SI{}{\nano\meter\squared}$), with total integration time of $\SI{41.6}{\second}$ and a single spectrum acquisition time of $\SI{10}{\milli\second}$. From each of the equivalent target sites we extract and sum 20 spectra to match the total integration time for single-shot EELS and use the same background subtraction to obtain the Cr $L_{3,2}$ edge (\textbf{Fig.~\ref{figEELSCSB}e}). We obtain the expected intensity, however, the measurement time to collect the same information makes targeted EELS $> 40$ times more dose efficient compared to collecting a full spectral image, dramatically reducing electron exposure to the specimen.


\section{6. Atomic Lock-On Precision for 1L WS$_2$}

\begin{figure*}[!ht]
	\scalebox{\figurescale}{\includegraphics[width=0.8\linewidth]{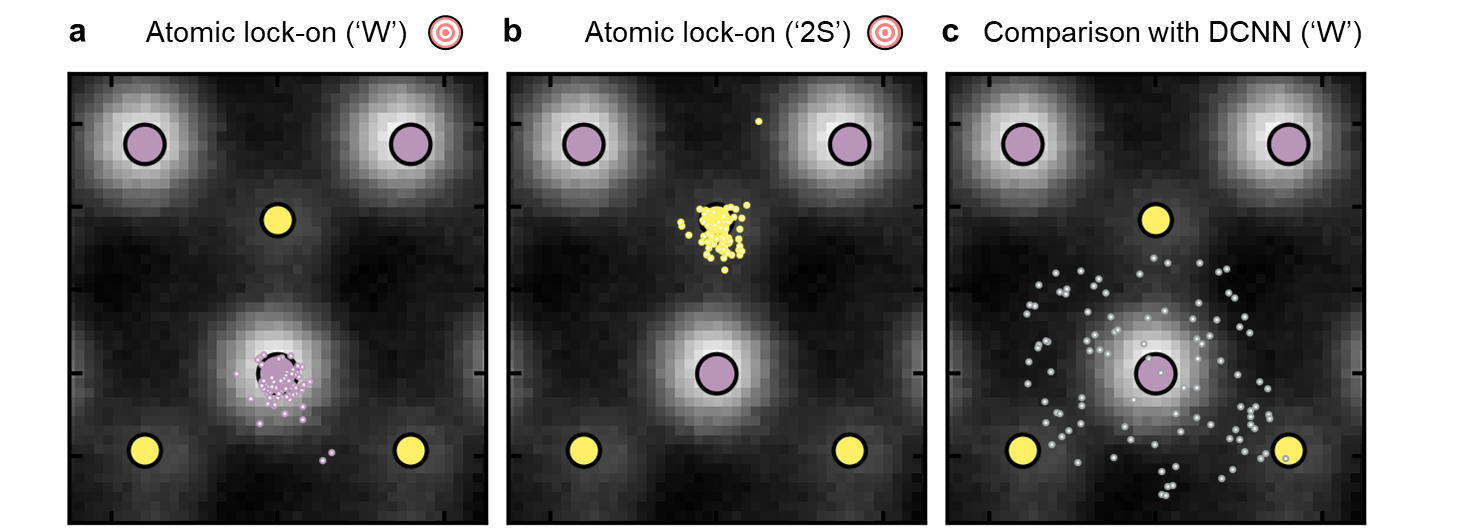}}
    \renewcommand{\figurename}{Figure}
	\caption{\label{fig_WS2}
		\textbf{ALO precision for targeting W and 2S atom columns on 1L WS$_2$.}
        \textbf{a, b}, ALO targeting of the W atom and 2S atom column with $28 \pm 18 \SI{}{\pico\meter}$ ad $27 \pm 17 \SI{}{\pico\meter}$ precision, respectively.
        \textbf{c}, Comparison to targeting the W atom column using a DCNN.
		}
\end{figure*}

\newpage

\section{7. Additional Single-Atom Time Dynamics.}

\begin{figure*}[!ht]
	\scalebox{\figurescale}{\includegraphics[width=1\linewidth]{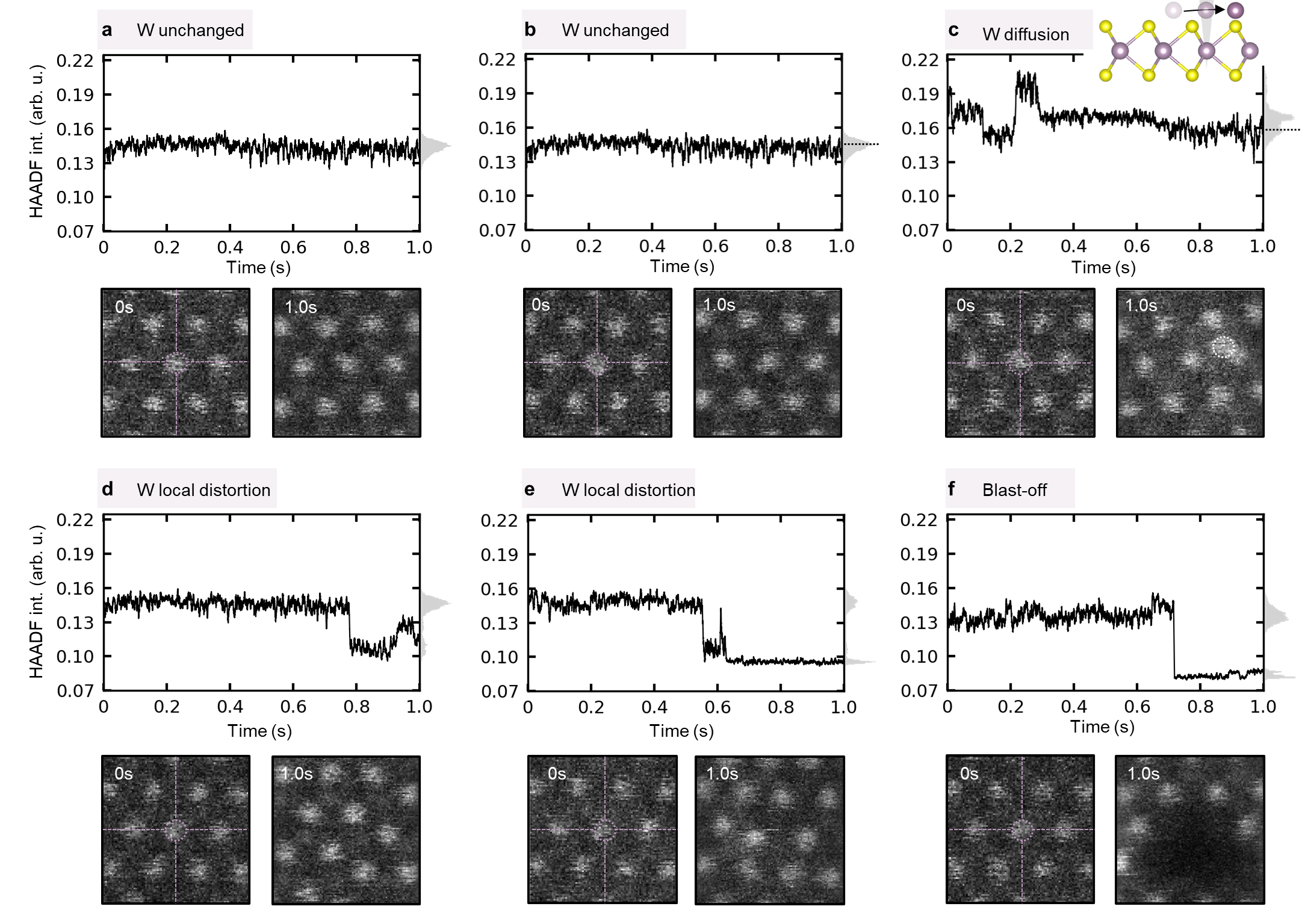}}
    \renewcommand{\figurename}{Figure}
	\caption{\label{fig_W_time}
		\textbf{Single-atom dynamics targeting W in 1L WS$_2$.}
		\textbf{a, b}, Unchanged crystal structure after 1s of targeting W atom column.
  \textbf{c}, Adatom diffusion with a HAADF intensity in excellent agreement with a W atom. In the final frame after the exposure a W atom is visible in close proximity.
  \textbf{d, e}, Beam induced local lattice distortions result in in-plane shifts of the W atom out of the electron beam. 
  \textbf{f}, Beam induced material removal ('blast off') creating a nanopore. A beam energy of $\SI{60}{\kilo\electronvolt}$ and a beam current of $\SI{20}{\pico\ampere}$ was used corresponding to $2.5 \cdot 10^8 e^- \SI{}{\per\angstrom\squared} s^{-1} $.
		}
\end{figure*}

\newpage

\begin{figure*}[!ht]
	\scalebox{\figurescale}{\includegraphics[width=1\linewidth]{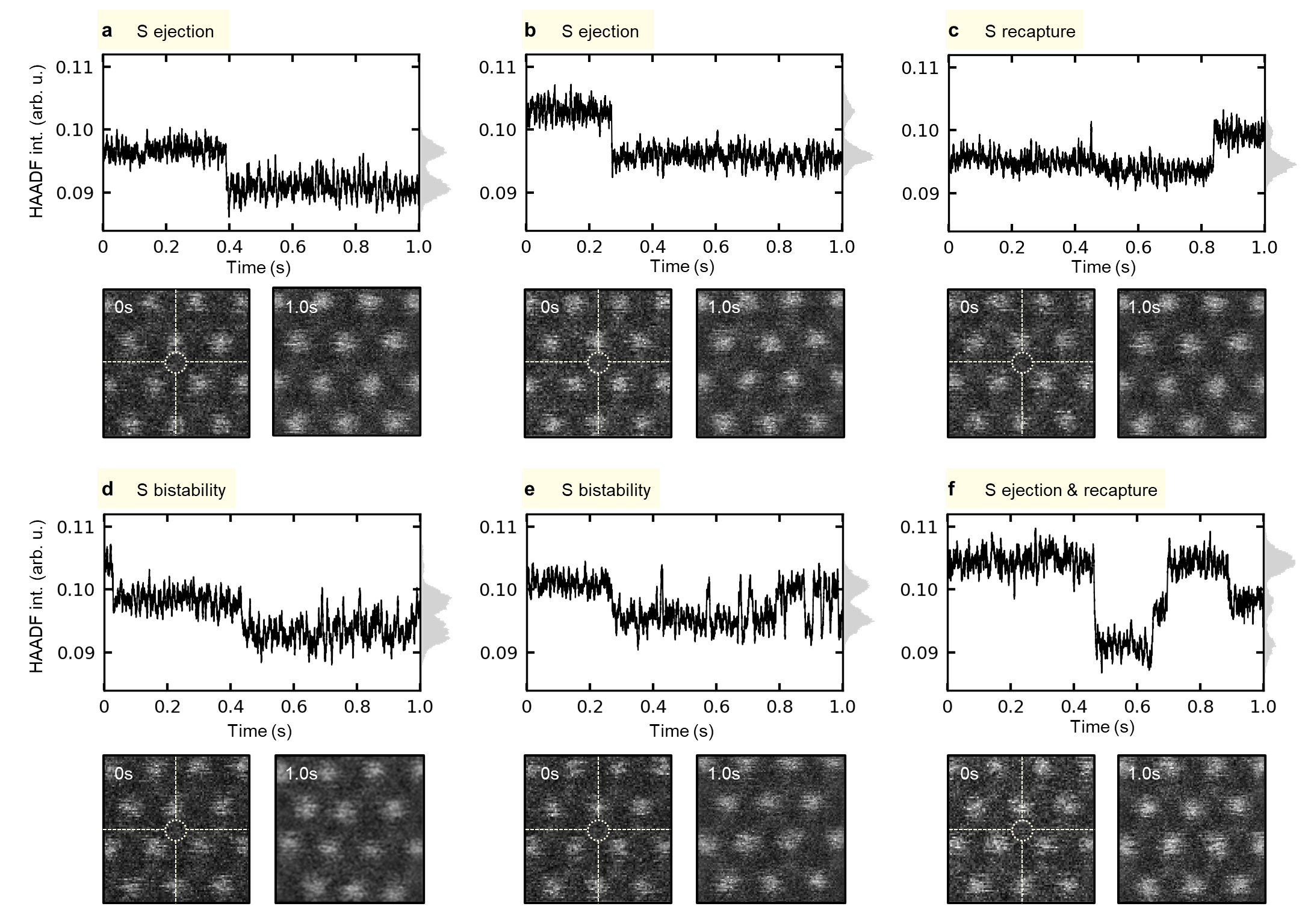}}
    \renewcommand{\figurename}{Figure}
	\caption{\label{fig_S_time}
		\textbf{Single-atom dynamics targeting 2S in 1L WS$_2$.}
		\textbf{a, b}, Ejection of a S atom.
  \textbf{c}, Recapture of a S atom.
  \textbf{d, e} Random telegraph noise as a signature of bistability of a S atom moving between the exposed site and a proximal position. 
  \textbf{f}, Ejection and recapture of S atoms. A beam energy of $\SI{60}{\kilo\electronvolt}$ and a beam current of $\SI{20}{\pico\ampere}$ was used corresponding to $2.5 \cdot 10^8 e^- \SI{}{\per\angstrom\squared} s^{-1} $.
		}
\end{figure*}

\newpage

\newpage



%
%
%
%
\bibliographystyle{apsrev}
\bibliography{full}